\documentclass[superscriptaddress,twocolumn,showpacs,footinbib,amssymb,aps]{revtex4}

\usepackage{graphicx,dcolumn,bm,amsmath,color,bm}

\newcommand{\eq}[1]{Eq.~(\ref{#1})}

\newcommand{\fig}[1]{Fig.~\ref{#1}}

\newcommand{\eeq}{ \end{equation} }
\newcommand{\beq}{ \begin{equation} }

\newcommand{\eea}{ \end{align} }
\newcommand{\bea}{ \begin{align} }

\newcommand{\bhu}{ {\bf \hat{u}} }

\newcommand{\bn}{ {\bf \hat{n}} }
\newcommand{\kbt}{k_{B}T}
\newcommand{\ellp}{\ell^{\prime}}
\newcommand{\ellpp}{\ell^{\prime \prime}}

\newcommand{\talpha}{\tilde{\alpha}}

\begin{document}

\title{Spinodal instabilities in polydisperse lyotropic nematics }
\author{C. Ferreiro-C\'ordova and H. H. Wensink}
\email{wensink@lps.u-psud.fr}
\affiliation{Laboratoire de Physique des Solides - UMR 8502, CNRS, Universit\'e Paris-Sud, Universit\'{e} Paris-Saclay, 91405 Orsay, France}

\date{\today}

\begin{abstract}

Many lyotropic liquid crystals are composed of mesogens that display a considerable spread in size or shape affecting their material properties and thermodynamics via various demixing and multi-phase coexistence scenarios.
Starting from a generalized Onsager theory we formulate a generic framework that enables locating spinodal polydispersities as well as identifying the nature of incipient size fractionation for arbitrary model  potentials and size distributions. We apply our theory to nematic phases of both hard rods and disks whose main particle dimension is described by a unimodal log-normal distribution. We find that both rod-based and discotic nematics become unstable at a critical polydispersity of about 20 \%.  We also investigate the effect of doping nematic assemblies with a small fraction of large species and highlight their effect on the stability of the uniform nematic  fluid. We find that while rod-based are only weakly affected by the presence of large species, doping discotic nematics with very large platelets leads to a remarkable suppression of the spinodal instabilities.  This could open up routes towards controlling the mechanical properties of nematic materials by manipulating the local stability of nematic fluid and its tendency to undergo fractionation-driven microphase separation. 

\end{abstract}

\maketitle
\section{Introduction}

Polydispersity is ubiquitous in colloidal and polymeric systems since their constituents are hardly ever fully identical but exhibit a continuous spread in size, shape or charge.
A disparity in microscopic interactions can have a considerable impact on the phase stability \cite{huang_1991} as well as on the mechanical properties of colloidal and nanoscale materials via aggregation \cite{bushell_jcis1998}, packing \cite{farr_jcp2009} and percolation processes  \cite{kyrylyuk_PNAS2008}. Research efforts can be aimed at either purifying colloidal suspensions in order to achieve colloidal crystallization \cite{murray_kagan}, e.g. via templating \cite{blaaderen_wiltzius1997}, or at purposefully enhancing their size polydispersity, e.g. to improve the electronic conductivity of percolated rod networks \cite{kyrylyuk_NatNANO2011,kyrylyuk_PNAS2008}, to stabilize vitrified states of matter in spherical systems \cite{auer_nature2001,pusey_2009} or to tailor the rheological properties of complex fluids \cite{tenbrinke_sm2007,tenbrinke_sm2008}. 

Anisotropic colloids with a distinct  rod- or disk-shape seem to be particularly prone to forming polydisperse mixtures whether the synthesis procedure is controlled or of a natural origin \cite{davidson-overview,davidson1997}. Basic examples range from clay suspension composed of thin sheets with variable diameter \cite{paineau2013},  mineral rods with strong length polydispersity  \cite{buining_jacs1991,vdpol_jcp2008,woolston_jcp2015}, variable-length filamentous biopolymers such as cellulose nanocrystals \cite{lagerwall2014a} or actin \cite{furukawa_actin1993}, to polydisperse carbon-based nanotubes (CNTs) \cite{baughman_cnt2002} and graphene oxide sheets \cite{xu_graphene_LC_2011}.  

While often overlooked for fluids of spherical particles \cite{pusey_nature1986, schaertl1994,bellier_jcp2000}, the effect of size disparity on the liquid crystal phase behaviour of anisotropic particles  has been the subject of considerable theoretical scrutiny (see Refs. \cite{vroege92, wensinkthesis, mederos_overview2014} and references therein). In order to ease the computational burden, most theoretical models considered thus far were restricted to simple binary mixtures of two different sizes or shapes \cite{lekkerkerker_coulon84,stroobants1984,vroij_mulder1998,purdy_prl2005,wensink_vroegeJPCM2004,martinez-ratonJCP2005,phillips_pre2010,shundyak_pre2003,bier_pre2004}, or ternary mixtures comprising three components \cite{vroege_lekkerkerker1997}. 
Truly polydisperse systems, however, involve a sheer infinite number of particles which makes the computation of phase diagrams a highly non-trivial task both in theory \cite{sollichwarrenmomenttheory,sollichoverview} and computer simulation \cite{Kofke99,wilding_jcp2002}.  One of the main complications of these systems  is that their multi-component character enables coexistence of arbitrarily many fluid or solid phases each accommodating a specific subpopulation of the overall distribution. A prominent example are dense fluids of polydisperse spherical particles where the presence of a range of different sizes may lead to a suppression of crystallization \cite{bartlett_prl1999, phan_jcp1998} in favour of various amorphous states whose nature has been the subject of recent  simulation studies  \cite{yianno_jcp2009, sarkar_pre2013,vdlinden_jcp2013}. Alternatively, size polydisperse sphere fluids may develop a coexistence of a number of different solid phases following a pathway referred to as fractionated crystallization \cite{bartlett_jcp1998}. This scenario which involves several solid phases in coexistence, each containing a much narrower distribution of particle sizes than is present in the system overall, was initially borne out by numerical calculations \cite{fasolo_prl2003,sollich_prl2010,botet_2016} and corroborated by recent experimental findings \cite{cabane_prl2016}. 

Experimental studies of synthetic clays reveal that liquid crystals composed of strongly polydisperse anisotropic constituents too are prone to forming multiple coexisting phases driven by a repartitioning of the various particles sizes and shapes. Columnar phases of plate-shaped gibbsite colloids can withstand a significant degree of diameter polydispersity \cite{kooij_nature2000} but  fractionated multiple columnar states may appear upon prolonged standing \cite{byelov_2010}. Similar observations of polydispersity-driven nematic-nematic phase separation have been reported for sepiolite rods \cite{zhang_jcp2006}. Similarly,  nematic liquid crystals formed in suspensions of beidellite platelets may develop a clear interface over time pointing at several fractioned phases in coexistence \cite{paineau_jpcb2009,pc1}.

While many factors may be at play in determining the complicated interactions between clay particles and their phase behaviour, such as particle flexibility, surface charge patterns depending on ionic strength and pH, gravity, etcetera, it is commonly assumed that their vast size polydispersity is chiefly responsible for the observed demixing. The key questions we set out to answer are the following: Can we find evidence for a critical polydispersity -- defined as the variance in rod length or disk diameter -- needed to drive a polydisperse nematic fluid into demixing? What is the role of particle shape, i.e., how do the spinodal instabilities compare for rod and plate-based nematics? And last but not least:  How is the stability of the nematic phase affected by doping a unimodal parent system with a small fraction of particles representing the tail populations of the size distribution, i.e. very large species?  

In order to address these problems we reconsider Onsager's classic second-virial theory, suitably extended to treat fully polydisperse lyotropic nematics \cite{speranza_pre2003,wensink_jcp2003}, and propose a tractable route to calculating the critical size distribution needed to render homogeneous nematic fluids unstable with respect to demixing.   Rather than embarking on full phase-split calculations such as done for the thermotropic case within a similar but simplified Onsager-type approach \cite{sollich_jcp2005}, our purpose here is to focus on spinodal instabilities of both rod- and disk-based lyotropic nematics. These spinodal instabilities provide insight into the {\em initial} stages of phase separation including the associated fractionation behaviour eventually leading up to a macroscopic phase split comprising two or more coexisting nematic phases. The advantages of taking this approach are twofold; first the theory remains numerically manageable and, second, we retain a generic framework that can be applied to arbitrary parent distributions and particle potentials and shapes, provided the microscopic parameters are such as to warrant the global stability of nematic order with respect to more ordered liquid crystal structures.  We first apply our formalism to the cases of infinitely long hard rods and infinitely thin hard disks with a unimodal distribution of their main particle dimension. We then look into the effect of doping these unimodal distribution with a small quantity of large species and scrutinize its impact on the critical polydispersity as well as the fractionation mechanism dominating the initial stages of phase separation.

\section{Theory}
In the following, we will adhere to the discussion and notation used in Ref. \cite{wensink_jcp2003}.
 Let us start with the free energy per unit volume $V$ of a polydisperse assembly of strongly anisometric hard particles (rods or disks) whose primary dimension (rod length $L$ or disk diameter $D$) is described by some quenched distribution $c(\ell)$. Onsager theory \cite{Onsager} states that in the second-virial approximation: 
\begin{align}
f = \frac{v_{0}F}{\kbt V} & \sim \int d \ell c(\ell) (\ln c(\ell) - 1) + \int d \ell c(l) \sigma (l)  \nonumber \\ 
& + \iint d \ell  d \ellp c(\ell) \rho(\ell , \ellp) c(\ellp) 
\label{free}
\end{align}
where $\ell$ denotes the main particle dimension, which could be either the length of a slender rod or the diameter of a thin disk,  exhibiting a continuous spread prescribed by a normalised distribution $p(\ell)$, so that $c(\ell) = c_{0} p(\ell)$ in terms of the overall particle density $c_{0} = Nv_{0}/V $ and microscopic volume $v_{0}$ which we will specify later on. 
 The polydisperse dimension  $\ell$ is normalized with respect to its {\em average} value, so that $\ell = L/\langle L \rangle$ for rods and $\ell = D/\langle D \rangle$ for platelets.
 Here, bare brackets define averages  (moments) of  the size distribution via $\langle \cdot \rangle = \int_{0}^{\infty} d \ell p(\ell)(\cdot)$.

The three entropic contributions in \eq{free} relate to the ideal gas, orientational, and excluded volume entropy, respectively. These quantities are defined as weighted averages of the normalised orientational distribution function $\psi(\theta, \ell)$ of each species. The angle $\theta$ represents the polar angle between the main particle orientation vector $\bhu$ and preferred direction of alignment $\bn$, so that $\cos \theta = \bhu \cdot \bn$. Assuming strongly ordered nematics, we may use a Gaussian Ansatz  $\psi (\theta, \ell) \sim \alpha(\ell) \exp [ -\frac{1}{2} \alpha(\ell) \theta^{2}  ]/4 \pi $ with its polar mirror form $\psi (\pi - \theta , \ell)$ to express the entropic contributions in terms of a single variational parameter $\alpha(\ell)$ quantifying the degree of nematic order. For asymptotically large $\alpha$ it is possible to obtain a tractable expression for the orientational entropy: 
\begin{align}
\sigma(\ell) & = \langle \ln [ 4 \pi \psi ( \Omega, \ell ) ] \rangle_{\psi}  \sim  \ln \alpha(\ell) - 1,
\label{sigma}
\end{align}
The packing entropy $\rho$ is defined as the excluded volume {\em per particle} $v_{ex}$ in  the nematic with respect to its respective value in the isotropic phase where the particles point in random directions.  We now identify the microscopic volume $v_{0}$ with the isotropic excluded volume per particle of {\em average-sized} species. Then:
\beq
\rho(\ell, \ellp)  \equiv \frac{v_{ex}}{v_{0}} =  \tilde{v}(\ell , \ellp)  \frac{\langle \langle  | (\bhu \times  \bhu^{\prime} |  \rangle \rangle_{\psi}}{\langle \langle  | (\bhu \times  \bhu^{\prime} |  \rangle \rangle_{\psi_{iso}}}
\label{rho}
\eeq
where brackets denote orientational averages $\langle \cdot \rangle_{\psi} = \int d \Omega \psi(\Omega, \ell) (\cdot)$ in terms of a solid angle $\Omega$. Noting that in the isotropic phase $\psi_{iso} =(4 \pi)^{-1}$ one easily finds that $\langle \langle  | (\bhu \times  \bhu^{\prime} |  \rangle \rangle_{\psi_{iso}}  \equiv \frac{\pi}{4}$. 
For sufficiently anisotropic particle shapes (slender rods and thin disks) the isotropized excluded volumes are $v_{0} \equiv  v_{ex}^{iso} = \frac{\pi}{4} \langle L \rangle^{2} D$ for rods and $v_{0} = \frac{\pi^{2}}{16} \langle D \rangle^{3}$ for platelets. An asymptotic expression for the nematic part in \eq{rho} can be obtained using the Gaussian trial functions defined earlier. For strongly ordered nematics $\alpha(\ell) \gg1 $ ($\forall \ell$) one obtains \footnote{Strictly, the asymptotic theory does not apply to very short particles ($\ell \ll 1)$ which have a tendency to adopt weakly nematic or isotropic order when embedded in a matrix of strongly aligned large particles.}: 
\beq
\langle \langle  | (\bhu \times  \bhu^{\prime} |  \rangle \rangle_{\psi}  \sim   \left ( \frac{\pi}{2}  \right )^{\frac{1}{2}}  \left ( \frac{1}{\alpha(\ell)} + \frac{1}{\alpha(\ellp)}  \right )^{\frac{1}{2}}
\label{singam}
\eeq
The dimensionless prefactor $\tilde{v}$ in \eq{rho} relates to the shape-dependent weight functions pertaining to hard cylindrical particles with extreme shape anisotropy. 
\beq
\tilde{v}  (\ell , \ellp)  = 
\begin{cases}  \ell \ellp &   \mbox{{\small rods:} }   \frac{D} {\langle L \rangle} \downarrow 0  \\ 
\ell \ellp  (\ell + \ellp)/2  &    \mbox{{\small disks:} }  \frac{L}{\langle D \rangle} \downarrow 0 \\ 
 \end{cases}
\eeq
Putting the entropic contributions \eq{sigma} and \eq{rho} back into the expression for the free energy $f$ and functional minimization with respect to the variational function $\alpha(\ell)$ yields an analytically insoluble integral equation for $\alpha$:
\beq
\talpha^{\frac{1}{2}}(\ell) =2^{\frac{1}{2}} \int d \ellp \tilde{v} (\ell , \ellp) p(\ellp) g_{0}(\ell, \ellp)
\label{alfa}
\eeq
with:
\beq
g_{0}(\ell, \ellp) = \left  ( 1 + \frac{\talpha(\ell)}{\talpha(\ellp)} \right )^{-\frac{1}{2}}
\eeq
Regardless of the particle size distribution, the degree of nematic alignment scales quadratically with density, so that it is expedient to factorize $\alpha (\ell) = 4 c_{0}^{2} \talpha (\ell)/\pi$ with the renormalised value $\talpha(\ell)$ depending only on the {\em shape} of the distribution $p(\ell)$ of particle dimensions in the nematic phase.  By reinserting \eq{alfa} into \eq{free} and elaborate rearranging, Odijk \cite{odijk1986} has shown that the  excess free energy given by the last contribution in \eq{free} equals a constant, namely 
$f_{\rm ex} / c_{0} = \langle \langle c_{0} \rho (\ell, \ellp) \rangle \rangle  = 2$. This remarkable outcome of the Gaussian variational approach holds true for any symmetric function $v_{\rm ex} (\ell , \ellp)$  and size distribution. Consequently, the osmotic pressure $P$ of the nematic phase is independent of particle composition and scales linearly with the total particle concentration, since \cite{odijk1986}:
\beq
\frac{Pv_{0}}{\kbt} =  c_{0} + c_{0}^{2} \frac{\partial (f_{\rm ex}/c_{0})}{\partial c_{0}} \sim 3 c_{0}
\label{pressure}
\eeq
Given that the excluded volume entropy is insensitive to $p(\ell)$, a possible nematic-nematic demixing instability must be driven by a competition between mixing and orientational entropy alone \cite{OdijkLekkerkerker,vroege1993}.   In view of \eq{pressure}  any two coexisting nematic phases must have the same overall density $c_{0}$. 
The chemical potential, defined as  $\mu(\ell) = \kbt \delta f / \delta c(\ell) $, depends essentially on the shape of the parent distribution with the concentration serving as an irrelevant off-set:
\begin{align}
\beta \mu(\ell ) &= {\rm cst} + 3 \log c_{0} + \ln [ \talpha (\ell) p(\ell) ] \nonumber \\
&+ 2^{\frac{3}{2}} \int d \ellp \tilde{v} (\ell , \ellp) p(\ellp)  h_{0}(\ell, \ellp)
\label{mu}
\end{align}
where $h_{0}$ depends implicitly on $\talpha$ via (cf. \eq{rho}):
\beq
h_{0}(\ell ,\ellp ) =  \left ( \frac{1}{\talpha(\ell)} + \frac{1}{\talpha(\ellp)}  \right )^{\frac{1}{2}} 
\eeq

Let us now attempt to probe instabilities of the homogeneous nematic phase with some prescribed size distribution against small fluctuations in particle composition with amplitude $\varepsilon$. Let us consider a functional perturbation $p(\ell) + \varepsilon \delta p(\ell)$. Marginal stability of the homogeneous nematic is guaranteed if the free energy change associated with any compositional fluctuation $\delta p(\ell)$ is positive:
\beq
\delta^{2} f = \iint d \ell d \ellp \delta p(\ell)  \frac{\delta \mu (\ell) }{\delta p(\ellp ) }  \delta p (\ellp) > 0
\eeq
for all species $\ell$. The functional derivative of the chemical potential can be obtained by noting that:
\beq
\delta \mu(\ell) = \int d \ellp \left ( \frac{\delta \mu(\ell)}{\delta p(\ellp)} \right ) \delta p(\ellp) 
\eeq
The explicit expression for the perturbation of the chemical potential can be obtained from linearizing \eq{mu} with respect to $\delta p$. Straightforward rearranging leads to:
\begin{widetext}
\begin{align}
 \delta \mu(\ell) & \sim \int d \ellp \delta (\ell - \ellp)  \left [ \frac{1}{\talpha(\ellp)}\frac{\delta \talpha(\ellp)}{\delta p(\ellp)}   + \frac{1}{p(\ellp)} \right ] \delta p(\ellp) \nonumber \\
&+ 2^{\frac{3}{2}} \int d \ellp \tilde{v} (\ell , \ellp) \left [  h_{0}(\ell, \ellp) + p(\ellp) \frac{\partial h_{0}(\ell, \ellp)}{\partial \talpha(\ell)}\frac{\delta \talpha(\ell)}{\delta p(\ellp)} +  p(\ellp) \frac{\partial h_{0}(\ell , \ellp)}{\partial \talpha(\ellp)} \frac{ \delta \talpha(\ellp)}{\delta p(\ellp)} \right ] \delta p(\ellp)
\label{dmu}
\end{align}
\end{widetext}
From which the functional derivative is readily identified.  At the spinodal point there is an incipient fluctuation $\delta p^{\ast}(\ell)$ along which the free energy does not change.
The condition $\delta^{2}f=0$ translates into the following criterion for spinodal instability in a polydisperse system:
\beq
\delta p (\ell) = \lambda \int d \ellp K(\ell, \ellp) \delta p (\ellp)
\label{fredholm}
\eeq
which is a functional eigenvalue equation, also referred to as a homogeneous Fredholm equation of the second kind \cite{numrecipes}.  The kernel $K$ is defined as:
\begin{align}
K(\ell, \ellp ) &= -\frac{p(\ell)}{\talpha(\ellp)}  \frac{\delta \talpha(\ellp)}{\delta p(\ellp)}\delta (\ell - \ellp)  - 2^{\frac{3}{2}} \tilde{v}(\ell, \ellp) p(\ell) h_{0}(\ell, \ellp) \nonumber \\ 
 & \times \left [ 1 - \frac{p(\ellp)}{2} \left (  h_{1} (\ellp, \ell) \frac{\delta \talpha(\ell)}{\delta p (\ellp)}  + h_{1}(\ell , \ellp) \frac{\delta \talpha(\ellp)}{\delta p (\ellp)}   \right )   \right ] 
\end{align}
While $\delta p^{\ast}(\ell) =0$ is a trivial solution of \eq{fredholm}, our task is to find non-trivial  eigen functions   $\delta p^{\ast}(\ell) \neq 0$ corresponding to the eigenvalue $\lambda=1$. These eigenfunctions signal the onset of  nematic-nematic phase separation and the emergence of a new  nematic phase characterized by a size distribution different from that  of the parent phase.


The functional derivative of $ \talpha $ featuring in $K$ can be obtained from the self-consistency equation \eq{alfa} by straightforward functional differentiation.  Similar as before we use:
\beq
\delta \talpha(\ell) = \int d \ellp \left ( \frac{\delta \talpha(\ell) }{\delta p (\ellp)} \right ) \delta p(\ellp)  
\eeq
Tedious rearranging then leads to:
\beq
\frac{\delta \talpha (\ell)}{\delta p(\ellp)} = \frac{2^{\frac{3}{2}}  \tilde{v}(\ell, \ellp ) \talpha^{\frac{1}{2}}(\ell) \left [ g_{0}(\ell, \ellp) + p(\ellp) \frac{\partial g_{0}(\ell, \ellp)}{\partial \talpha (\ellp)}  \frac{\delta \talpha (\ellp)}{\delta p (\ellp)}  \right ]  } {1 - 2^{\frac{3}{2}} \tilde{v}(\ell , \ellp) p(\ellp) \talpha^{\frac{1}{2}}(\ell) \frac{\partial g_{0}(\ell, \ellp)}{\partial \talpha (\ell)} } 
\eeq
Setting $\ell = \ellp$ and noting that $g_{0}(\ell, \ell ) = 2^{-\frac{1}{2}}$ we get the simple diagonal term:
\beq
 \frac{\delta \talpha (\ell)}{\delta p (\ell)}  = 2 \talpha^{\frac{1}{2}}(\ell)  \tilde{v} (\ell, \ell) 
 \label{diag1}
 \eeq
Generally, integral equations such as \eq{fredholm} can only be solved analytically if the kernel $K$ is separable,  which is clearly not the case here. Some insight can be gained from an approximate solution of \eq{alfa} by considering an infinitely narrow size distribution described by a delta-distribution $\delta(\ell - 1)$.  The solution of \eq{alfa} then turns out:
\beq
\talpha(\ell) \sim \frac{1}{2} \tilde{v}(1,1)^{2} \left \{ \left ( 1+ 8 \frac{\tilde{v}^{2}(\ell, 1) }{\tilde{v}^{2}(1,1) } \right )^{\frac{1}{2}}  -1  \right \}
\eeq
Taking the large-size limit  ($\ell \gg 1$) leads to $\talpha (\ell) \propto \tilde{v}(\ell, 1)$ revealing that $\talpha$ scales linearly with $\ell$ for rods and quadratically with $\ell$ for disks.  For finite-width parent distributions one must resort to numerical routes to solve the governing equations.  A numerical solution of the self-consistency equation \eq{alfa} and the eigenvalue problem  \eq{fredholm} is easily obtained by discretizing the kernel on an equidistant grid $ \ell_{i} \in  [ \ell_{\rm min} , \ell_{\rm max} ] $ and using standard Simpson's numerical quadrature integration and diagonalization packages from {\em Mathematica} \textregistered. 

\begin{figure}
\begin{center}
\includegraphics[width= \columnwidth]{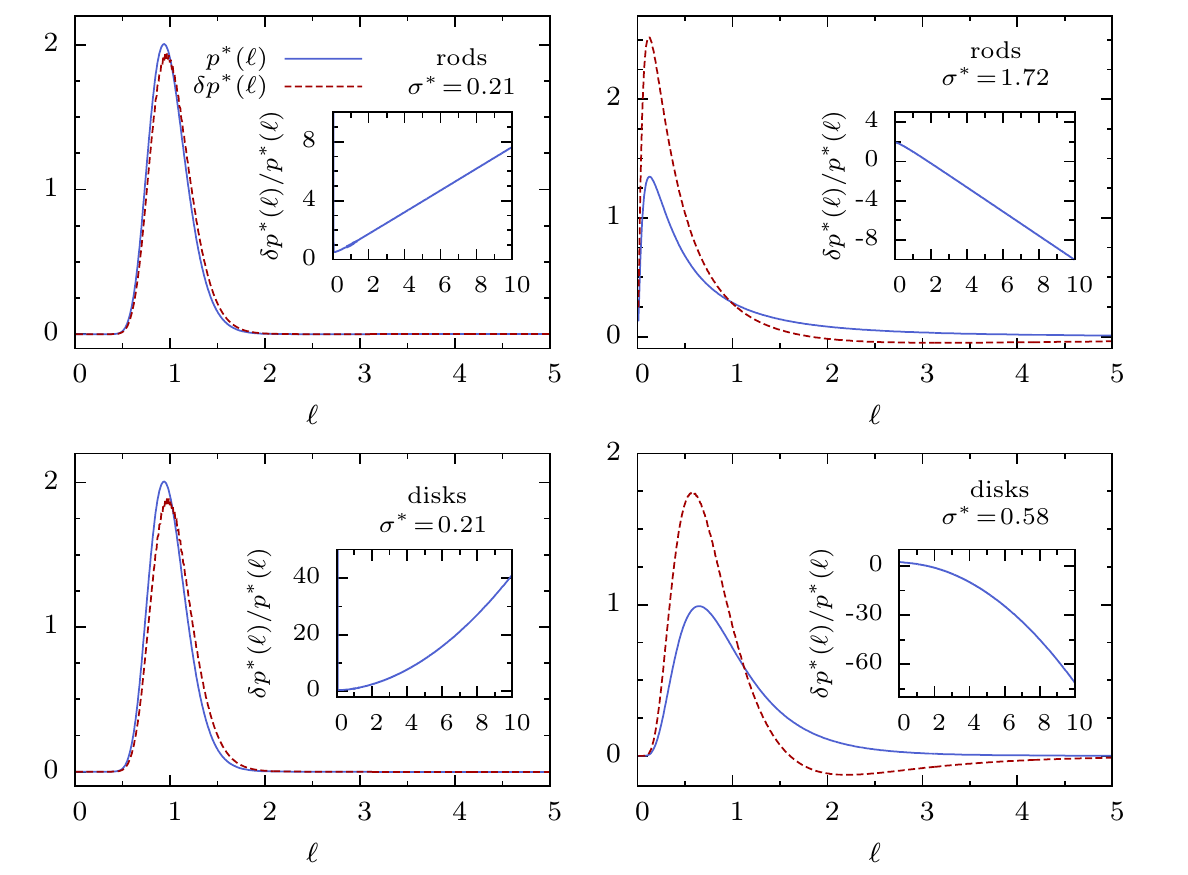}
\caption{ \label{fig1} Overview of the parent distributions $p(\ell)$ associated with the two nematic-nematic spinodal points, and the corresponding eigenfunctions $\delta p^{\ast}(\ell)$ for a log-normal distribution with cut-off values $\ell_{\rm min} = 0.01$ and $\ell_{\rm max} = 10$.  The left panels correspond to low-$\sigma$ spinodal instabilities where  the small species are fractionated out into the new phase, whereas the inverse happens at the high-$\sigma$ spinodal point (right panels). }
\end{center}
\end{figure}

\begin{figure}
\begin{center}
\includegraphics[width= \columnwidth]{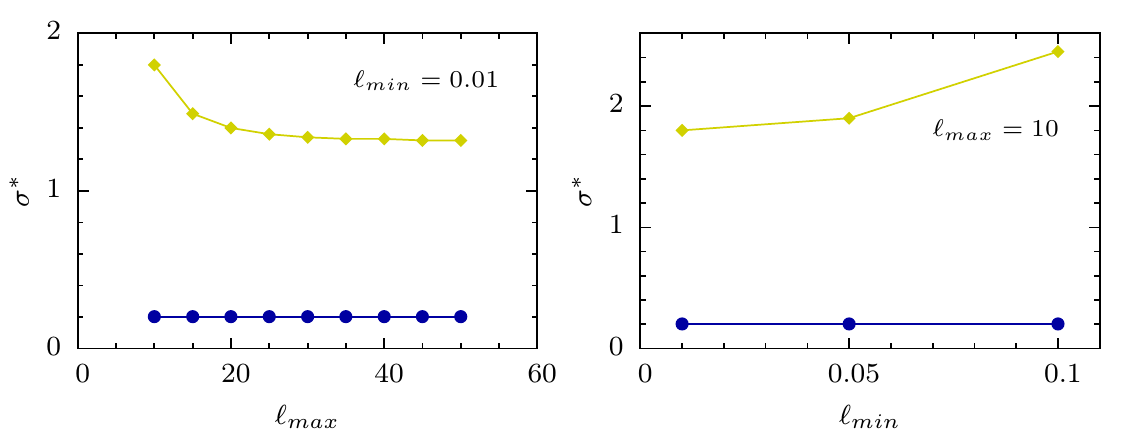}
\caption{ \label{fig2} Shift of the spinodal points upon changing the upper ($\ell_{\rm max}$) and lower ($\ell_{\rm min}$) cut-off values for hard {\em rods} with a log-normal length distribution.  The trends for hard disks (not shown) turn out to be marginal. }
\end{center}
\end{figure}

\section{Results for unimodal size distribution}

An appropriate form for the typical size distribution of colloidal particles with quenched polydispersity is the log-normal contribution which has the following form:
\beq
p(\ell) = \frac{ 1}{(2 \pi )^{\frac{1}{2}} w \ell } \exp \left [ -\frac{( \ln \ell + \frac{w^{2}}{2} )^{2}}{2 w^{2}}  \right ]
\label{ln}
\eeq
with bounds $\ell_{\rm min} =0 $ and $\ell_{\rm max} \rightarrow \infty $. \eq{ln} has mean $\langle \ell \rangle = 1$  and size polydispersity $\sigma$ is connected to the standard deviation via $\sigma^{2} = e^{w^{2}} -1$. Finite-tail cutoffs lead to small corrections for which we account numerically.  Results for the spinodal instabilities occurring in a polydisperse nematic system of rods or disks with a  log-normal size distribution have been compiled in \fig{fig1}. 

\begin{figure*}
\begin{center}
\includegraphics[width= 1.9 \columnwidth]{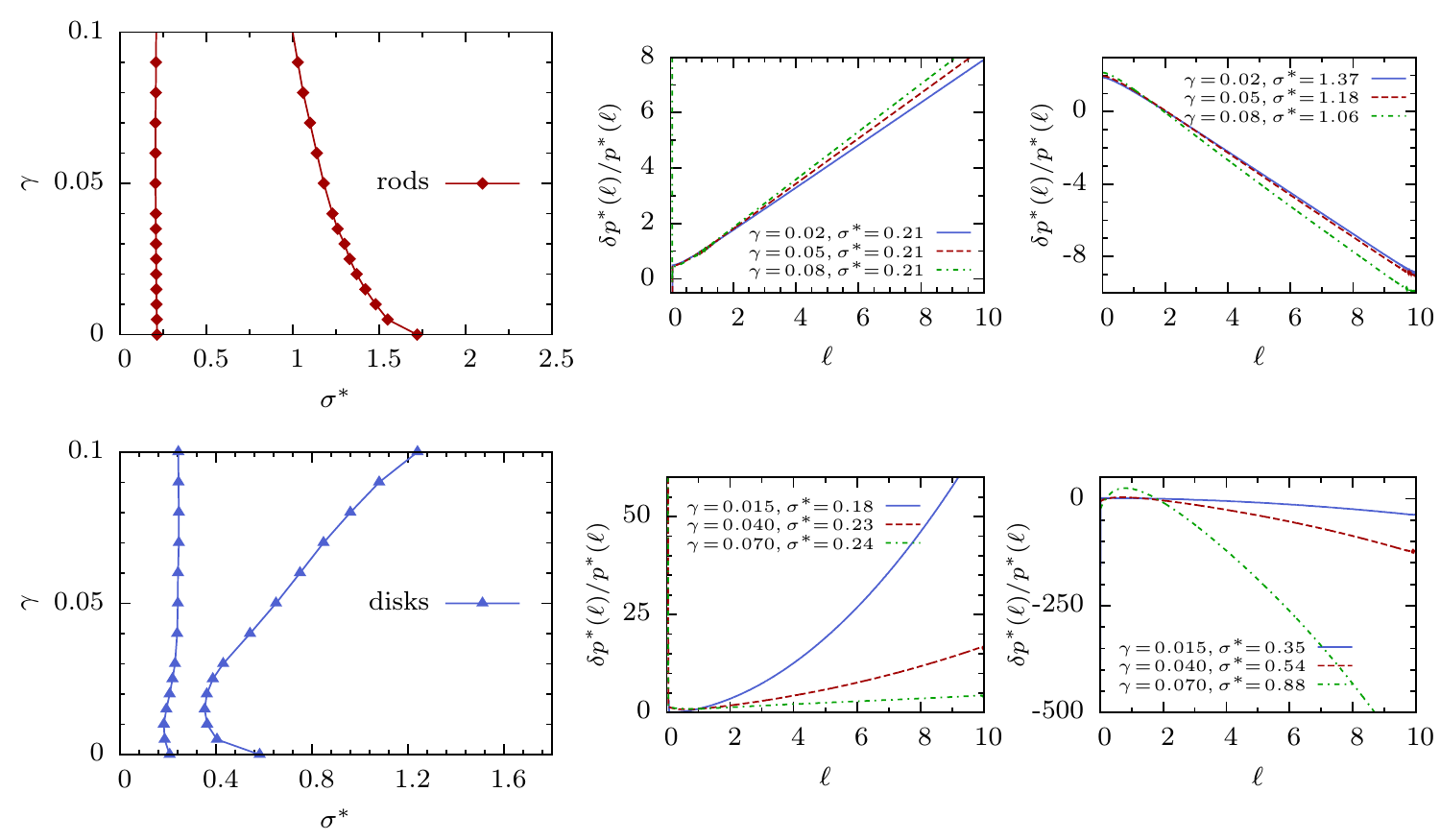}
\caption{ \label{fig3} (left panels) Effect of large species dopant on the spinodal polydispersities $\sigma^{\ast}$ for both rod- and disk-type nematics with log-normal size distributions $p_{d}(\ell)$ with bounds $\ell_{\rm min} = 0.01$ and $\ell_{\rm max} = 10$ and $\ell_{0} = 1$. The parameter $\gamma$ roughly corresponds to the mole fraction of largest species.  The right panels demonstrate how size fractionation $\delta p^{\ast}(\ell)/p^{\ast}(\ell)$ is affected by the fraction of dopant.  For discotic nematics adding a few percent of dopant suppresses fractionation at the lower $\sigma^{\ast}$ but reinforces fractionation at the high-$\sigma$ spinodal.    }
\end{center}
\end{figure*}

We observe that both cases display a lower spinodal point occurring at a polydispersity of $\sigma =0.21$, a value that is by all means a realistic one given that many mineral clay particles have typical polydispersity beyond that value. A second spinodal point appears at a much higher polydispersity than the first with the value for rods being much larger than the one for plates. The eigenfunctions associated with each of the spinodals identify a fractionation scenario whereby the longest species are split off from the mother distributions for the low-$\sigma$ one whereas the opposite happens at the second spinodal. There is a marked difference between the degree of fractionation which clearly is much stronger for the platelets than for the rods. Since the results shown in \fig{fig1} only hold for a particular choice of cut-off values, we ought to verify how sensitive they are upon changing the range of particle dimensions considered. It turns out that they are robust. This is certainly true for the platelets where a change of lower or upper cut-off has a marginal effect; for the rods cut-off effects are more pronounced though still quite weak as shown in  \fig{fig2}.  In the Appendix we provide supplementary results on the local curvature of the free energy in the vicinity of the spinodal. This is to ensure that the spinodals we have been identifying actually correspond to local {\em minima} in the free energy landscape of a polydisperse mixture rather than local maxima in which case the spinodal instabilities do not have a physically relevant meaning. The general picture emerging from \fig{fig4} shown in the Appendix is that the species occupying the tails of the distribution  ($\ell \sim \ell_{\rm max}$) correspond to the sections with the largest negative curvature and therefore constitute the main driving force of the demixing. In line with the fractionation strength in \fig{fig1} we observe a stark contrast in the amplitudes of $\delta^{3} f$ with the disk showing a far greater propensity to demix than the rods. 

An alternative distribution that is commonly used to represent polymer molecular weight distributions is the Schulz-Zimm  function \cite{schulz1939,zimm1948}:
\beq
p(\ell) = \frac{(1+z)^{1+z}}{\Gamma (1+z)} \ell^{z} \exp (-(z + 1) \ell )
\eeq
As  \eq{ln} it is normalized on the domain $0< \ell < \infty$ and has mean $\langle \ell \rangle = 1$ and polydispersity $\sigma = (1+z)^{-1/2}$. Finite value cut-off values are accounted for numerically. The
lower spinodal instabilities for the Schulz-Zimm distribution turn out to be virtually identical to those of the log-normal one ($\sigma \approx 0.2$ for both rods and disks). The large-$\sigma$ values differ somewhat   ($\sigma \approx 1.46 $ for the rods and $\sigma \approx 0.65$ for the disks) in view of the different relative weight of the tail parts of the distributions.

\section{Effect of large-species dopant}

In this section we address the question as to what happens with the spinodals if a unimodally distributed parent mixture is doped with a tiny fraction of very large species. To this end we supplement the log-normal distribution with a growing exponential tail to construct a weakly bimodal size distribution:
\beq
p_{d}(\ell) \sim p(\ell) + \gamma 
\begin{cases}  e^{a ( \ell_{\rm max} - \ell)} &  \ell \ge (\ell_{\rm max} -\ell_{0}) \\ 
0  & \ell < (\ell_{\rm max}   - \ell_{0})  \\ 
 \end{cases}
 \label{pd}
\eeq
 with $\gamma \ll 1$ representing the mole fraction of largest species and $a = \ell_{0}^{-1} \ln (p(\ell_{\rm max} -\ell_{0})/\gamma)$ ensuring a continuous connection between the two branches and $ \ell_{0} $ governing the size spread of the dopant. \eq{pd} lacks a trivial normalization factor which is included in the numerical calculations. An overview of the effect of the dopant on the stability of a homogenous polydisperse nematic phase is given in \fig{fig3} for both rods and disks. The presence of dopant clearly affects the trends of the spinodals even at very small mole fractions. For both particle shapes the high-$\sigma$ spinodal appears to be affected the most by the presence of the large species. For discotic nematics significant non-monotonic trends can be discerned in which the distance between the two spinodals first reduces significantly, reaching a minimum at about 1 to 2 percent mass fraction of dopant, before widening up considerably at higher mole fractions. The associated fractionation scenarios are depicted on the right in \fig{fig3}. For rod-based nematic the impact of adding a tiny fraction of macroscopically sized rods seems fairly marginal with the fractionation strength being enhanced somewhat. The case of disks however, reveal a much richer scenario. The most striking feature occurs at the low-$\sigma$ spinodal; while fractionation is reinforced at very small $\gamma \sim 0.015$ (leading to slight reduction of the critical polydispersity $\sigma^{\ast}$),  a strong reduction in fractionation is observed at larger dopant concentration ($\gamma \sim 0.04$). In fact, the corresponding free energy curvature  $\delta^{3}f$ depicted in \fig{fig5} in the Appendix turns out to be uniformly positive indicating that the spinodal no longer pinpoints a local free energy minimum but rather a maximum. This suggest the possibility of a {\em complete suppression} of spinodal demixing in polydisperse discotic nematics by adding a few percent of dopant consisting of macroscopically large plate-shaped objects. 
  
\section{Conclusions}

In this paper we have analyzed spinodal instabilities and the possibility of nematic-nematic demixing in nematic phases composed of strongly polydisperse hard rod- or disk-shaped particles. To this end we have used used Onsager's classical second virial theory extended to mixtures with a continuous distribution of particle sizes. Our motivation to revisit the problem of size polydispersity on liquid crystal stability stems from recent observations of nematic-nematic demixing in a number of mineral clay systems, such as sepiolite rods \cite{zhang_jcp2006} and beidellite platelets \cite{paineau_jpcb2009}.   

We have found that both rod and plate-based nematics develop spinodal instabilities at typical polydispersities of 20 \%. This result is rather indensitive to the precise details of the size distribution provided the function is smooth and unimodal (cf. log-normal or Schulz-Zimm). The spinodals are hallmarks of a fluid-fluid phase separation where the system ultimately develops two or more coexisting nematic phases each harboring  a different portion of the overall size distribution. This critical polydispersity is an experimentally relevant one given that most anisotropic mineral colloids capable of forming liquid crystals are characterized by (much) larger size polydispersities. This suggests that lyotropic nematic phases of strongly anisometric building blocks have the tendency to fluid-fluid phase separate (either mesoscopically through some fractionation-mediated microphase separation or via a full macroscopic phase split) at much weaker polydispersities than anticipated.

In our study we have also pointed out that doping nematic assemblies with small fraction of very large particles allows for a careful control of the location of the spinodal, most notably for discotic lyotropic nematics in which case the propensity to undergo spinodal demixing can be completely suppressed by adding a few percent of macroscopically large disk-shaped objects.  This unusual scenario could be exploited to tune the properties of nematic materials by manipulating the local stability of the nematic fluid, in particular the extent to which it is subject to fractionation-driven microphase-separation. 

Naturally, with our static theory we are not in a position to make any statements about the kinetics of phase separation and, more particularly, on the time-scales associated with spinodal demixing. Based on experimental oberservations in clay systems mentioned before we expect a macroscopic nematic-nematic demixing to be a very slow process spanning a time-scale of months or even years \cite{pc1}. The results of our calculations highlight the presence of spinodal instabilities and suggest that many nematic liquid crystalline systems composed of polydisperse building blocks may, despite their apparent stability, develop signs of fractionation-driven phase separation over time. 

An interesting future challenge would be to develop a dynamic density functional theory that allows for a detailed monitoring of the initial stages of spinodal demixing \cite{dhontbook,dhontbrielsPRE2005}.  Within this framework one could trace the time-evolution of the size distribution of the incipient fractionated nematic phase at the early stages of demixing where the microstructure of the nematic is not yet dominated by many-particle hydrodynamic interactions (which are notoriously difficult to account for theoretically).  However, in view of the ultraslow and quasi-adiabatic nature of the nematic-nematic demixing dynamics one might be capable of reliably tracing the demixing kinetics up to much larger time scales without the need to account for these hydrodynamics modes. Studying these type of fractionation processes could provide an interesting opportunity for the application of conventional dynamical density functional theories whose adiabatic description relies entirely on particle-particle correlations for a fluid at thermal equilibrium \cite{marconi1999,archer2004,rex2007}.

\section*{Acknowledgments}

The authors acknowledge helpful discussions with Robert Botet and Patrick Davidson. This work was funded by a Young Researchers (JCJC) grant from the French National Research Agency (ANR). 

\section{Appendix: curvature of the free energy landscape}

\begin{figure}
\begin{center}
\includegraphics[width=0.9 \columnwidth]{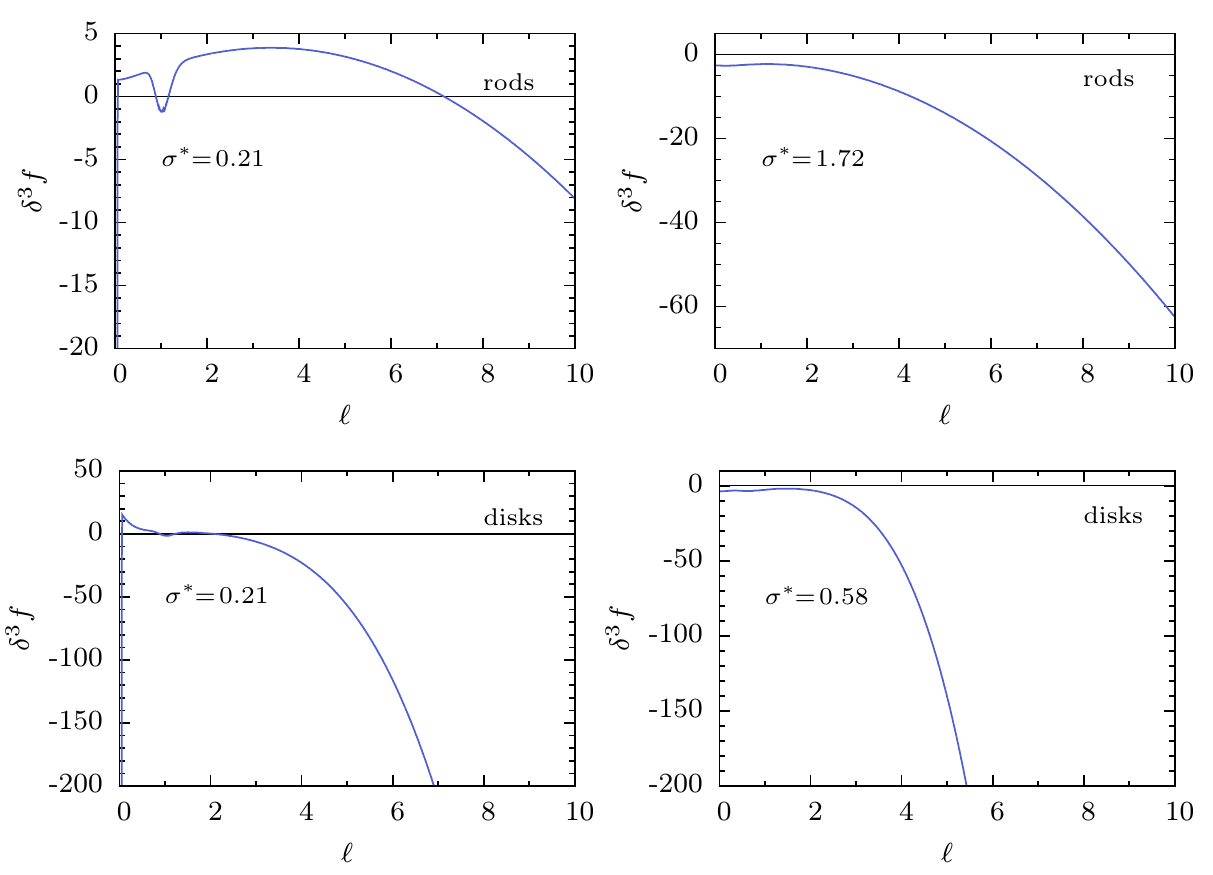}
\caption{ \label{fig4}  Free energy curvature $\delta^{3} f$ indicating whether the spinodal inflection points correspond to a local minimum (if negative) or local maximum.   }
\end{center}
\end{figure}

\begin{figure}
\begin{center}
\includegraphics[width=0.9 \columnwidth]{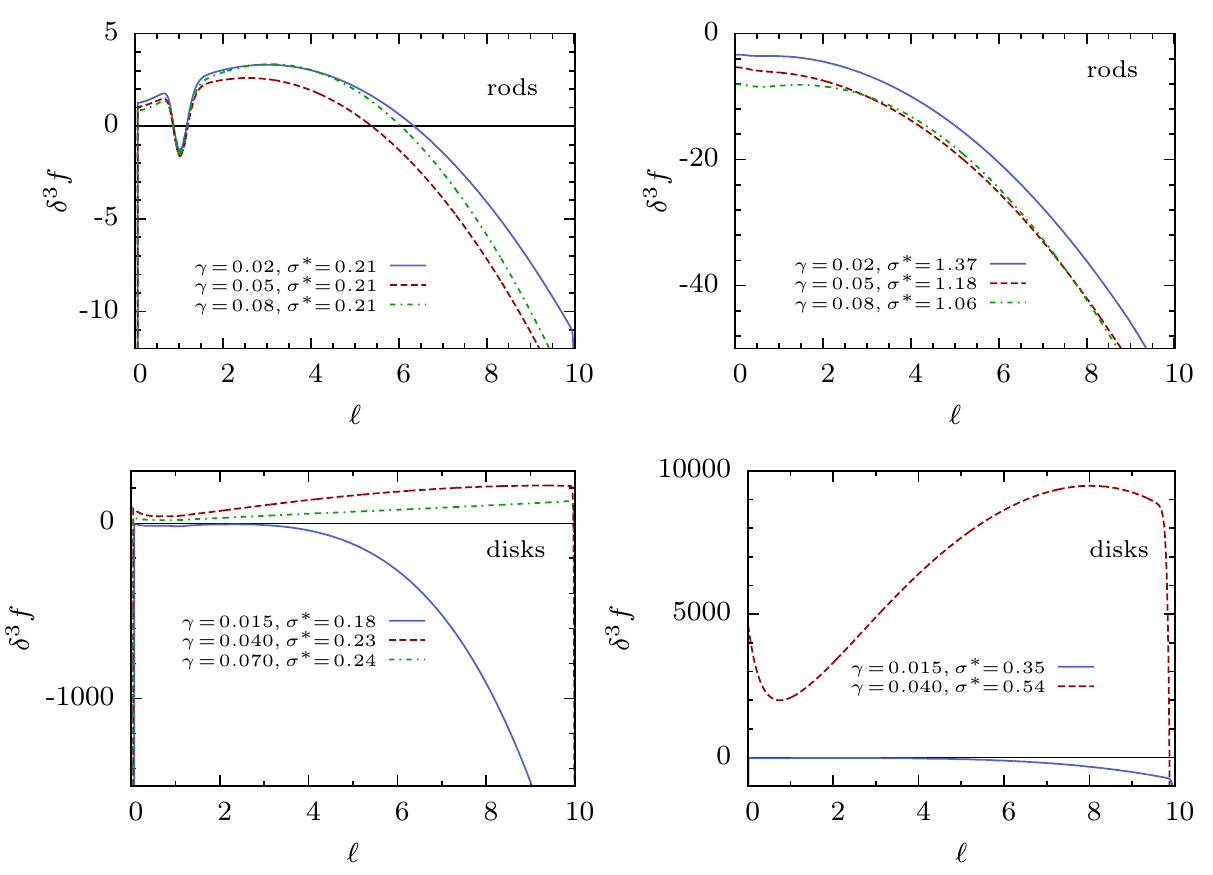}
\caption{ \label{fig5}  Free energy curvature $\delta^{3} f$ for the {\em large-species doped} systems at different dopant mole fractions $\gamma$.  }
\end{center}
\end{figure}

In order to ascertain whether the spinodal instability signals a local minimum or a local maximum in the free energy landscape we need to compute the third derivative of the free energy functional \eq{free}. Functional Taylor expansion of the chemical potential \eq{mu} formally gives:
\begin{align}
& \mu[p(\ell) + \delta p(\ell)] = \mu [p(\ell ) ] +  \int d \ellp \left ( \frac{\delta \mu(\ell)}{\delta p(\ellp)} \right ) \delta p(\ellp) \nonumber \\ 
& + \frac{1}{2} \iint d \ellp d \ellpp \left ( \frac{\delta^{2} \mu(\ell)}{\delta p(\ellp) \delta p(\ellpp)} \right ) \delta p(\ellp) \delta p (\ellpp) + \cdots 
\end{align}
At the spinodal point $\delta p(\ell) = \delta p^{\ast}(\ell )$ (with $p^{\ast}$ the corresponding parent distribution) the second term drops out as per \eq{fredholm} and the local curvature direction of the free energy landscape follows from:
\beq
\delta^{3} f(\ell)  =
\frac{1}{2} \iint d \ellp d \ellpp   \left ( \frac{\delta^{2} \mu(\ell)}{\delta p(\ellp) \delta p(\ellpp)} \right ) _{p = p ^{\ast}} \delta p^{\ast}(\ellp) \delta p^{\ast} (\ellpp) 
\eeq
such that $\delta ^{3} f(\ell) < 0 $ denotes a local minimum leading to a nematic-nematic demixing, whereas $ \delta^{3} f(\ell) > 0$ corresponds to a physically irrelevant local maximum, and $\delta^{3} f(\ell) =0$ to a critical point.  The second-order derivate of the chemical potential can be derived along the lines of \eq{dmu}. Lengthy derivations then lead to 
\begin{widetext}
\begin{align}
\delta^{3} f & =\frac{1}{2} \left [ \frac{1}{\talpha}  \frac{\delta^{2} \talpha}{\delta p^{2}}  - \frac{1}{\talpha^{2}} \left ( \frac{\delta \talpha}{\delta p } \right )^{2} - \frac{1}{p^{2}} \right ]_{p=p^{\ast}} (\delta p^{\ast})^{2} \nonumber \\ 
& + 2^{\frac{1}{2}}  \int d \ellp  \tilde{v}(\ell , \ellp) \left \{  \frac{\partial h_{0}}{\partial \talpha}  \frac{\delta \talpha }{\delta p'}  +  \frac{\partial h_{0}}{\partial \talpha'}  \frac{\delta \talpha' }{\delta p'}   - \frac{1}{2} h_{01} \left ( \frac{\delta \talpha }{\delta p'} + p'   \frac{\delta^{2} \talpha}{\delta p'^{2}} \right ) - \frac{1}{2} h_{10} \left ( \frac{\delta \talpha' }{\delta p'}  + p'  \frac{\delta^{2} \talpha'}{\delta p'^{2}} \right ) \right  . \nonumber \\
& \left .  -\frac{p'}{2} \left ( \frac{\partial h_{01}}{\partial \talpha} \frac{\delta \talpha}{\delta p'}  +  \frac{\partial h_{01}}{\partial \talpha'} \frac{\delta \talpha'}{\delta p'} \right )  \frac{\delta \talpha}{\delta p'}
-\frac{p'}{2} \left ( \frac{\partial h_{10}}{\partial \talpha} \frac{\delta \talpha}{\delta p'}  +  \frac{\partial h_{10}}{\partial \talpha'} \frac{\delta \talpha'}{\delta p'} \right )  \frac{\delta \talpha'}{\delta p'} \right \}_{p=p^{\ast}} (\delta p'^{\ast})^{2}
 \label{f3}
 \end{align}
\end{widetext}
where we have used short-hand notations  $p = p(\ell)$, $p' = p(\ellp)$, etcetera. Furthermore $h_{01} = h_{0}(\ell, \ellp) h_{1}(\ellp, \ell)$ and $h_{10} = h_{0}(\ell, \ellp)h_{1}(\ell, \ellp)$.
The second derivative of the nematic order parameter $\talpha$ is given by:
\begin{widetext}
\begin{align}
\frac{\delta^{2} \talpha}{\delta p'^{2}}=\frac{2^{\frac{3}{2}}\tilde{v} \left \{ \frac{\partial \talpha^{\frac{1}{2}} g_{0}}{\partial \talpha} \frac{\delta \talpha}{\delta p'} + \frac{\partial \talpha^{\frac{1}{2}} g_{0}}{\partial \talpha'} \frac{\delta \talpha'}{\delta p'}       +       \talpha^{\frac{1}{2}} \frac{\partial  g_{0}}{\partial \talpha} \frac{\delta \talpha}{\delta p'} + \talpha^{\frac{1}{2}} \frac{\partial  g_{0}}{\partial \talpha'} \frac{\delta \talpha'}{\delta p'}  +  {p' \talpha^{\frac{1}{2}} \frac{\partial g_{0}}{\partial \talpha '}  \frac{\delta^{2}\talpha ' }{\delta p'^{2}}} + W
\right \}  }{1-  2^{\frac{3}{2}}\tilde{v} p' \talpha^{\frac{1}{2}} \frac{\partial g_{0}}{\partial \talpha}
}
 \end{align}
\end{widetext}
where $W$ stands for:
\begin{widetext}
\begin{align}
W &=p' \frac{\partial}{\partial \talpha} \left ( \talpha^{\frac{1}{2}} \frac{\partial g_{0}}{\partial \talpha} \right ) \frac{\delta \talpha}{\delta p'}  \frac{\delta \talpha}{\delta p'} +
p' \frac{\partial}{\partial \talpha'} \left ( \talpha^{\frac{1}{2}} \frac{\partial g_{0}}{\partial \talpha} \right ) \frac{\delta \talpha'}{\delta p'}  \frac{\delta \talpha}{\delta p'} +
p' \frac{\partial}{\partial \talpha} \left ( \talpha^{\frac{1}{2}} \frac{\partial g_{0}}{\partial \talpha'} \right ) \frac{\delta \talpha}{\delta p'}  \frac{\delta \talpha'}{\delta p'}  + p' \frac{\partial}{\partial \talpha'} \left ( \talpha^{\frac{1}{2}} \frac{\partial g_{0}}{\partial \talpha'} \right ) \frac{\delta \talpha'}{\delta p'}  \frac{\delta \talpha'}{\delta p'}
\end{align}
\end{widetext}
The diagonal term reads:
\beq
\frac{\delta^{2}\talpha (\ell)}{\delta p(\ell)^{2}} = 2 \tilde{v}(\ell, \ell)^{2} 
\eeq
This term naturally also follows from taking the derivative of \eq{diag1}. Some representative curves are shown in \fig{fig4} and \fig{fig5} for the undoped and doped distributions, respectively.  While $\delta^{3}f$ is uniformly negative for the disks, the results for the rods reveal some intervals where $\delta^{3}f >0$ indicating unstable spinodal modes. For the discotics doping brings about a drastic change in the free energy landscape whereby the lower spinodal $\sigma^{\ast}$ identifies a local maximum instead of a minimum, thereby suppressing spinodal demixing.

\bibliography{pub}

\begin{thebibliography}{71}
\expandafter\ifx\csname natexlab\endcsname\relax\def\natexlab#1{#1}\fi
\expandafter\ifx\csname bibnamefont\endcsname\relax
  \def\bibnamefont#1{#1}\fi
\expandafter\ifx\csname bibfnamefont\endcsname\relax
  \def\bibfnamefont#1{#1}\fi
\expandafter\ifx\csname citenamefont\endcsname\relax
  \def\citenamefont#1{#1}\fi
\expandafter\ifx\csname url\endcsname\relax
  \def\url#1{\texttt{#1}}\fi
\expandafter\ifx\csname urlprefix\endcsname\relax\def\urlprefix{URL }\fi
\providecommand{\bibinfo}[2]{#2}
\providecommand{\eprint}[2][]{\url{#2}}

\bibitem[{\citenamefont{Huang and Radosz}({1991})}]{huang_1991}
\bibinfo{author}{\bibfnamefont{S.}~\bibnamefont{Huang}} \bibnamefont{and}
  \bibinfo{author}{\bibfnamefont{M.}~\bibnamefont{Radosz}},
  \bibinfo{journal}{{Ind. Eng. Chem. Res.}} \textbf{\bibinfo{volume}{{30}}},
  \bibinfo{pages}{{1994}} (\bibinfo{year}{{1991}}).

\bibitem[{\citenamefont{Bushell and Amal}({1998})}]{bushell_jcis1998}
\bibinfo{author}{\bibfnamefont{G.}~\bibnamefont{Bushell}} \bibnamefont{and}
  \bibinfo{author}{\bibfnamefont{R.}~\bibnamefont{Amal}}, \bibinfo{journal}{{J.
  Colloid Interface Sci.}} \textbf{\bibinfo{volume}{{205}}},
  \bibinfo{pages}{{459}} (\bibinfo{year}{{1998}}).

\bibitem[{\citenamefont{Farr and Groot}({2009})}]{farr_jcp2009}
\bibinfo{author}{\bibfnamefont{R.~S.} \bibnamefont{Farr}} \bibnamefont{and}
  \bibinfo{author}{\bibfnamefont{R.~D.} \bibnamefont{Groot}},
  \bibinfo{journal}{{J. Chem. Phys.}} \textbf{\bibinfo{volume}{{131}}},
  \bibinfo{pages}{{244104}} (\bibinfo{year}{{2009}}).

\bibitem[{\citenamefont{Kyrylyuk and van~der
  Schoot}({2008})}]{kyrylyuk_PNAS2008}
\bibinfo{author}{\bibfnamefont{A.~V.} \bibnamefont{Kyrylyuk}} \bibnamefont{and}
  \bibinfo{author}{\bibfnamefont{P.}~\bibnamefont{van~der Schoot}},
  \bibinfo{journal}{{Proc. Natl. Acad. Sci. U.S.A.}}
  \textbf{\bibinfo{volume}{{105}}}, \bibinfo{pages}{{8221}}
  (\bibinfo{year}{{2008}}).

\bibitem[{\citenamefont{Murray et~al.}({2000})\citenamefont{Murray, Kagan, and
  Bawendi}}]{murray_kagan}
\bibinfo{author}{\bibfnamefont{C.}~\bibnamefont{Murray}},
  \bibinfo{author}{\bibfnamefont{C.}~\bibnamefont{Kagan}}, \bibnamefont{and}
  \bibinfo{author}{\bibfnamefont{M.}~\bibnamefont{Bawendi}},
  \bibinfo{journal}{{Annu. Rev. Mater. Sci.}} \textbf{\bibinfo{volume}{{30}}},
  \bibinfo{pages}{{545}} (\bibinfo{year}{{2000}}).

\bibitem[{\citenamefont{van Blaaderen et~al.}({1997})\citenamefont{van
  Blaaderen, Ruel, and Wiltzius}}]{blaaderen_wiltzius1997}
\bibinfo{author}{\bibfnamefont{A.}~\bibnamefont{van Blaaderen}},
  \bibinfo{author}{\bibfnamefont{R.}~\bibnamefont{Ruel}}, \bibnamefont{and}
  \bibinfo{author}{\bibfnamefont{P.}~\bibnamefont{Wiltzius}},
  \bibinfo{journal}{{Nature}} \textbf{\bibinfo{volume}{{385}}},
  \bibinfo{pages}{{321}} (\bibinfo{year}{{1997}}).

\bibitem[{\citenamefont{Kyrylyuk et~al.}({2011})\citenamefont{Kyrylyuk,
  Hermant, Schilling, Klumperman, Koning, and van~der
  Schoot}}]{kyrylyuk_NatNANO2011}
\bibinfo{author}{\bibfnamefont{A.~V.} \bibnamefont{Kyrylyuk}},
  \bibinfo{author}{\bibfnamefont{M.~C.} \bibnamefont{Hermant}},
  \bibinfo{author}{\bibfnamefont{T.}~\bibnamefont{Schilling}},
  \bibinfo{author}{\bibfnamefont{B.}~\bibnamefont{Klumperman}},
  \bibinfo{author}{\bibfnamefont{C.~E.} \bibnamefont{Koning}},
  \bibnamefont{and} \bibinfo{author}{\bibfnamefont{P.}~\bibnamefont{van~der
  Schoot}}, \bibinfo{journal}{{Nat. Nanotechnol.}}
  \textbf{\bibinfo{volume}{{6}}}, \bibinfo{pages}{{364}}
  (\bibinfo{year}{{2011}}).

\bibitem[{\citenamefont{Auer and Frenkel}({2001})}]{auer_nature2001}
\bibinfo{author}{\bibfnamefont{S.}~\bibnamefont{Auer}} \bibnamefont{and}
  \bibinfo{author}{\bibfnamefont{D.}~\bibnamefont{Frenkel}},
  \bibinfo{journal}{{Nature}} \textbf{\bibinfo{volume}{{413}}},
  \bibinfo{pages}{{711}} (\bibinfo{year}{{2001}}).

\bibitem[{\citenamefont{Pusey et~al.}({2009})\citenamefont{Pusey, Zaccarelli,
  Valeriani, Sanz, Poon, and Cates}}]{pusey_2009}
\bibinfo{author}{\bibfnamefont{P.~N.} \bibnamefont{Pusey}},
  \bibinfo{author}{\bibfnamefont{E.}~\bibnamefont{Zaccarelli}},
  \bibinfo{author}{\bibfnamefont{C.}~\bibnamefont{Valeriani}},
  \bibinfo{author}{\bibfnamefont{E.}~\bibnamefont{Sanz}},
  \bibinfo{author}{\bibfnamefont{W.~C.~K.} \bibnamefont{Poon}},
  \bibnamefont{and} \bibinfo{author}{\bibfnamefont{M.~E.} \bibnamefont{Cates}},
  \bibinfo{journal}{{Philos. Trans. A Math. Phys. Eng. Sci.}}
  \textbf{\bibinfo{volume}{{367}}}, \bibinfo{pages}{{4993}}
  (\bibinfo{year}{{2009}}).

\bibitem[{\citenamefont{ten Brinke et~al.}({2007})\citenamefont{ten Brinke,
  Bailey, Lekkerkerker, and Maitland}}]{tenbrinke_sm2007}
\bibinfo{author}{\bibfnamefont{A.~J.~W.} \bibnamefont{ten Brinke}},
  \bibinfo{author}{\bibfnamefont{L.}~\bibnamefont{Bailey}},
  \bibinfo{author}{\bibfnamefont{H.~N.~W.} \bibnamefont{Lekkerkerker}},
  \bibnamefont{and} \bibinfo{author}{\bibfnamefont{G.~C.}
  \bibnamefont{Maitland}}, \bibinfo{journal}{{Soft Matter}}
  \textbf{\bibinfo{volume}{{3}}}, \bibinfo{pages}{{1145}}
  (\bibinfo{year}{{2007}}).

\bibitem[{\citenamefont{ten Brinke et~al.}({2008})\citenamefont{ten Brinke,
  Bailey, Lekkerkerker, and Maitland}}]{tenbrinke_sm2008}
\bibinfo{author}{\bibfnamefont{A.~J.~W.} \bibnamefont{ten Brinke}},
  \bibinfo{author}{\bibfnamefont{L.}~\bibnamefont{Bailey}},
  \bibinfo{author}{\bibfnamefont{H.~N.~W.} \bibnamefont{Lekkerkerker}},
  \bibnamefont{and} \bibinfo{author}{\bibfnamefont{G.~C.}
  \bibnamefont{Maitland}}, \bibinfo{journal}{{Soft Matter}}
  \textbf{\bibinfo{volume}{{4}}}, \bibinfo{pages}{{337}}
  (\bibinfo{year}{{2008}}).

\bibitem[{\citenamefont{Davidon and Gabriel}(2005)}]{davidson-overview}
\bibinfo{author}{\bibfnamefont{P.}~\bibnamefont{Davidon}} \bibnamefont{and}
  \bibinfo{author}{\bibfnamefont{J.~C.~P.} \bibnamefont{Gabriel}},
  \bibinfo{journal}{Curr. Opin. Colloid Interface Sci.}
  \textbf{\bibinfo{volume}{9}}, \bibinfo{pages}{377} (\bibinfo{year}{2005}).

\bibitem[{\citenamefont{Davidson et~al.}({1997})\citenamefont{Davidson, Batail,
  Gabriel, Livage, Sanchez, and Bourgaux}}]{davidson1997}
\bibinfo{author}{\bibfnamefont{P.}~\bibnamefont{Davidson}},
  \bibinfo{author}{\bibfnamefont{P.}~\bibnamefont{Batail}},
  \bibinfo{author}{\bibfnamefont{J.}~\bibnamefont{Gabriel}},
  \bibinfo{author}{\bibfnamefont{J.}~\bibnamefont{Livage}},
  \bibinfo{author}{\bibfnamefont{C.}~\bibnamefont{Sanchez}}, \bibnamefont{and}
  \bibinfo{author}{\bibfnamefont{C.}~\bibnamefont{Bourgaux}},
  \bibinfo{journal}{{Prog. Polym. Sci}} \textbf{\bibinfo{volume}{{22}}},
  \bibinfo{pages}{{913}} (\bibinfo{year}{{1997}}).

\bibitem[{\citenamefont{Paineau et~al.}({2013})\citenamefont{Paineau, Philippe,
  Antonova, Bihannic, Davidson, Dozov, Gabriel, Imperor-Clerc, Levitz, Meneau
  et~al.}}]{paineau2013}
\bibinfo{author}{\bibfnamefont{E.}~\bibnamefont{Paineau}},
  \bibinfo{author}{\bibfnamefont{A.~M.} \bibnamefont{Philippe}},
  \bibinfo{author}{\bibfnamefont{K.}~\bibnamefont{Antonova}},
  \bibinfo{author}{\bibfnamefont{I.}~\bibnamefont{Bihannic}},
  \bibinfo{author}{\bibfnamefont{P.}~\bibnamefont{Davidson}},
  \bibinfo{author}{\bibfnamefont{I.}~\bibnamefont{Dozov}},
  \bibinfo{author}{\bibfnamefont{J.~C.~P.} \bibnamefont{Gabriel}},
  \bibinfo{author}{\bibfnamefont{M.}~\bibnamefont{Imperor-Clerc}},
  \bibinfo{author}{\bibfnamefont{P.}~\bibnamefont{Levitz}},
  \bibinfo{author}{\bibfnamefont{F.}~\bibnamefont{Meneau}},
  \bibnamefont{et~al.}, \bibinfo{journal}{{Liq. Cryst. Rev.}}
  \textbf{\bibinfo{volume}{{1}}}, \bibinfo{pages}{{110}}
  (\bibinfo{year}{{2013}}).

\bibitem[{\citenamefont{Buining et~al.}({1991})\citenamefont{Buining,
  Pathmamanoharan, Jansen, and Lekkerkerker}}]{buining_jacs1991}
\bibinfo{author}{\bibfnamefont{P.~A.} \bibnamefont{Buining}},
  \bibinfo{author}{\bibfnamefont{C.}~\bibnamefont{Pathmamanoharan}},
  \bibinfo{author}{\bibfnamefont{J.~B.~H.} \bibnamefont{Jansen}},
  \bibnamefont{and} \bibinfo{author}{\bibfnamefont{H.~N.~W.}
  \bibnamefont{Lekkerkerker}}, \bibinfo{journal}{{J. Am. Ceram. Soc.}}
  \textbf{\bibinfo{volume}{{74}}}, \bibinfo{pages}{{1303}}
  (\bibinfo{year}{{1991}}).

\bibitem[{\citenamefont{van~den Pol et~al.}({2008})\citenamefont{van~den Pol,
  Thies-Weesie, Petukhov, Vroege, and Kvashnina}}]{vdpol_jcp2008}
\bibinfo{author}{\bibfnamefont{E.}~\bibnamefont{van~den Pol}},
  \bibinfo{author}{\bibfnamefont{D.~M.~E.} \bibnamefont{Thies-Weesie}},
  \bibinfo{author}{\bibfnamefont{A.~V.} \bibnamefont{Petukhov}},
  \bibinfo{author}{\bibfnamefont{G.~J.} \bibnamefont{Vroege}},
  \bibnamefont{and}
  \bibinfo{author}{\bibfnamefont{K.}~\bibnamefont{Kvashnina}},
  \bibinfo{journal}{{J. Chem. Phys.}} \textbf{\bibinfo{volume}{{129}}},
  \bibinfo{pages}{{164715}} (\bibinfo{year}{{2008}}).

\bibitem[{\citenamefont{Woolston and van
  Duijneveldt}({2015})}]{woolston_jcp2015}
\bibinfo{author}{\bibfnamefont{P.}~\bibnamefont{Woolston}} \bibnamefont{and}
  \bibinfo{author}{\bibfnamefont{J.~S.} \bibnamefont{van Duijneveldt}},
  \bibinfo{journal}{{J. Chem. Phys.}} \textbf{\bibinfo{volume}{{142}}},
  \bibinfo{pages}{{184901}} (\bibinfo{year}{{2015}}).

\bibitem[{\citenamefont{Lagerwall et~al.}(2014)\citenamefont{Lagerwall,
  Sch{\"u}tz, Salajkov\'{a}, Noh, Park, Scalia, and
  Bergstr{\"o}m}}]{lagerwall2014a}
\bibinfo{author}{\bibfnamefont{J.~P.~F.} \bibnamefont{Lagerwall}},
  \bibinfo{author}{\bibfnamefont{C.}~\bibnamefont{Sch{\"u}tz}},
  \bibinfo{author}{\bibfnamefont{M.}~\bibnamefont{Salajkov\'{a}}},
  \bibinfo{author}{\bibfnamefont{J.}~\bibnamefont{Noh}},
  \bibinfo{author}{\bibfnamefont{J.~H.} \bibnamefont{Park}},
  \bibinfo{author}{\bibfnamefont{G.}~\bibnamefont{Scalia}}, \bibnamefont{and}
  \bibinfo{author}{\bibfnamefont{L.}~\bibnamefont{Bergstr{\"o}m}},
  \bibinfo{journal}{NPG Asia Mater.} \textbf{\bibinfo{volume}{6}},
  \bibinfo{pages}{e80} (\bibinfo{year}{2014}).

\bibitem[{\citenamefont{Furukawa et~al.}({1993})\citenamefont{Furukawa, Kundra,
  and Fechheimer}}]{furukawa_actin1993}
\bibinfo{author}{\bibfnamefont{R.}~\bibnamefont{Furukawa}},
  \bibinfo{author}{\bibfnamefont{R.}~\bibnamefont{Kundra}}, \bibnamefont{and}
  \bibinfo{author}{\bibfnamefont{M.}~\bibnamefont{Fechheimer}},
  \bibinfo{journal}{{Biochemistry}} \textbf{\bibinfo{volume}{{32}}},
  \bibinfo{pages}{{12346}} (\bibinfo{year}{{1993}}).

\bibitem[{\citenamefont{Baughman et~al.}({2002})\citenamefont{Baughman,
  Zakhidov, and de~Heer}}]{baughman_cnt2002}
\bibinfo{author}{\bibfnamefont{R.}~\bibnamefont{Baughman}},
  \bibinfo{author}{\bibfnamefont{A.}~\bibnamefont{Zakhidov}}, \bibnamefont{and}
  \bibinfo{author}{\bibfnamefont{W.}~\bibnamefont{de~Heer}},
  \bibinfo{journal}{{Science}} \textbf{\bibinfo{volume}{{297}}},
  \bibinfo{pages}{{787}} (\bibinfo{year}{{2002}}).

\bibitem[{\citenamefont{Xu and Gao}({2011})}]{xu_graphene_LC_2011}
\bibinfo{author}{\bibfnamefont{Z.}~\bibnamefont{Xu}} \bibnamefont{and}
  \bibinfo{author}{\bibfnamefont{C.}~\bibnamefont{Gao}}, \bibinfo{journal}{{ACS
  Nano}} \textbf{\bibinfo{volume}{{5}}}, \bibinfo{pages}{{2908}}
  (\bibinfo{year}{{2011}}).

\bibitem[{\citenamefont{Pusey and van Megen}({1986})}]{pusey_nature1986}
\bibinfo{author}{\bibfnamefont{P.}~\bibnamefont{Pusey}} \bibnamefont{and}
  \bibinfo{author}{\bibfnamefont{W.}~\bibnamefont{van Megen}},
  \bibinfo{journal}{{Nature}} \textbf{\bibinfo{volume}{{320}}},
  \bibinfo{pages}{{340}} (\bibinfo{year}{{1986}}).

\bibitem[{\citenamefont{Schaertl and Sillescu}({1994})}]{schaertl1994}
\bibinfo{author}{\bibfnamefont{W.}~\bibnamefont{Schaertl}} \bibnamefont{and}
  \bibinfo{author}{\bibfnamefont{H.}~\bibnamefont{Sillescu}},
  \bibinfo{journal}{{J. Stat. Phys.}} \textbf{\bibinfo{volume}{{77}}},
  \bibinfo{pages}{{1007}} (\bibinfo{year}{{1994}}).

\bibitem[{\citenamefont{Bellier-Castella
  et~al.}({2000})\citenamefont{Bellier-Castella, Xu, and
  Baus}}]{bellier_jcp2000}
\bibinfo{author}{\bibfnamefont{L.}~\bibnamefont{Bellier-Castella}},
  \bibinfo{author}{\bibfnamefont{H.}~\bibnamefont{Xu}}, \bibnamefont{and}
  \bibinfo{author}{\bibfnamefont{M.}~\bibnamefont{Baus}}, \bibinfo{journal}{{J.
  Chem. Phys.}} \textbf{\bibinfo{volume}{{113}}}, \bibinfo{pages}{{8337}}
  (\bibinfo{year}{{2000}}).

\bibitem[{\citenamefont{Vroege and Lekkerkerker}(1992)}]{vroege92}
\bibinfo{author}{\bibfnamefont{G.~J.} \bibnamefont{Vroege}} \bibnamefont{and}
  \bibinfo{author}{\bibfnamefont{H.~N.~W.} \bibnamefont{Lekkerkerker}},
  \bibinfo{journal}{Rep. Prog. Phys.} \textbf{\bibinfo{volume}{55}},
  \bibinfo{pages}{1241} (\bibinfo{year}{1992}).

\bibitem[{\citenamefont{Wensink}(2004)}]{wensinkthesis}
\bibinfo{author}{\bibfnamefont{H.~H.} \bibnamefont{Wensink}}, Ph.D. thesis,
  \bibinfo{school}{Utrecht University} (\bibinfo{year}{2004}).

\bibitem[{\citenamefont{Mederos et~al.}({2014})\citenamefont{Mederos, Velasco,
  and Martinez-Raton}}]{mederos_overview2014}
\bibinfo{author}{\bibfnamefont{L.}~\bibnamefont{Mederos}},
  \bibinfo{author}{\bibfnamefont{E.}~\bibnamefont{Velasco}}, \bibnamefont{and}
  \bibinfo{author}{\bibfnamefont{Y.}~\bibnamefont{Martinez-Raton}},
  \bibinfo{journal}{{J. Phys.; Condens. Matter}}
  \textbf{\bibinfo{volume}{{26}}}, \bibinfo{pages}{{463101}}
  (\bibinfo{year}{{2014}}).

\bibitem[{\citenamefont{Lekkerkerker et~al.}(1984)\citenamefont{Lekkerkerker,
  Coulon, van~der Hagen, and Deblieck}}]{lekkerkerker_coulon84}
\bibinfo{author}{\bibfnamefont{H.~N.~W.} \bibnamefont{Lekkerkerker}},
  \bibinfo{author}{\bibfnamefont{P.}~\bibnamefont{Coulon}},
  \bibinfo{author}{\bibfnamefont{R.}~\bibnamefont{van~der Hagen}},
  \bibnamefont{and} \bibinfo{author}{\bibfnamefont{R.}~\bibnamefont{Deblieck}},
  \bibinfo{journal}{J. Chem. Phys.} \textbf{\bibinfo{volume}{80}},
  \bibinfo{pages}{3427} (\bibinfo{year}{1984}).

\bibitem[{\citenamefont{Stroobants and Lekkerkerker}(1984)}]{stroobants1984}
\bibinfo{author}{\bibfnamefont{A.}~\bibnamefont{Stroobants}} \bibnamefont{and}
  \bibinfo{author}{\bibfnamefont{H.~N.~W.} \bibnamefont{Lekkerkerker}},
  \bibinfo{journal}{J. Phys. Chem.} \textbf{\bibinfo{volume}{88}},
  \bibinfo{pages}{3669} (\bibinfo{year}{1984}).

\bibitem[{\citenamefont{van Roij et~al.}({1998})\citenamefont{van Roij, Mulder,
  and Dijkstra}}]{vroij_mulder1998}
\bibinfo{author}{\bibfnamefont{R.}~\bibnamefont{van Roij}},
  \bibinfo{author}{\bibfnamefont{B.}~\bibnamefont{Mulder}}, \bibnamefont{and}
  \bibinfo{author}{\bibfnamefont{M.}~\bibnamefont{Dijkstra}},
  \bibinfo{journal}{{Physica A}} \textbf{\bibinfo{volume}{{261}}},
  \bibinfo{pages}{{374}} (\bibinfo{year}{{1998}}).

\bibitem[{\citenamefont{Purdy et~al.}({2005})\citenamefont{Purdy, Varga,
  Galindo, Jackson, and Fraden}}]{purdy_prl2005}
\bibinfo{author}{\bibfnamefont{K.}~\bibnamefont{Purdy}},
  \bibinfo{author}{\bibfnamefont{S.}~\bibnamefont{Varga}},
  \bibinfo{author}{\bibfnamefont{A.}~\bibnamefont{Galindo}},
  \bibinfo{author}{\bibfnamefont{G.}~\bibnamefont{Jackson}}, \bibnamefont{and}
  \bibinfo{author}{\bibfnamefont{S.}~\bibnamefont{Fraden}},
  \bibinfo{journal}{{Phys. Rev. Lett.}} \textbf{\bibinfo{volume}{{94}}},
  \bibinfo{pages}{{057801}} (\bibinfo{year}{{2005}}).

\bibitem[{\citenamefont{Wensink and Vroege}({2004})}]{wensink_vroegeJPCM2004}
\bibinfo{author}{\bibfnamefont{H.~H.} \bibnamefont{Wensink}} \bibnamefont{and}
  \bibinfo{author}{\bibfnamefont{G.~J.} \bibnamefont{Vroege}},
  \bibinfo{journal}{{J. Phys. Condens. Matter}}
  \textbf{\bibinfo{volume}{{16}}}, \bibinfo{pages}{{S2015}}
  (\bibinfo{year}{{2004}}).

\bibitem[{\citenamefont{Martinez-Raton
  et~al.}({2005})\citenamefont{Martinez-Raton, Velasco, and
  Mederos}}]{martinez-ratonJCP2005}
\bibinfo{author}{\bibfnamefont{Y.}~\bibnamefont{Martinez-Raton}},
  \bibinfo{author}{\bibfnamefont{E.}~\bibnamefont{Velasco}}, \bibnamefont{and}
  \bibinfo{author}{\bibfnamefont{L.}~\bibnamefont{Mederos}},
  \bibinfo{journal}{{J. Chem. Phys.}} \textbf{\bibinfo{volume}{{123}}},
  \bibinfo{pages}{{104906}} (\bibinfo{year}{{2005}}).

\bibitem[{\citenamefont{Phillips and Schmidt}({2010})}]{phillips_pre2010}
\bibinfo{author}{\bibfnamefont{J.}~\bibnamefont{Phillips}} \bibnamefont{and}
  \bibinfo{author}{\bibfnamefont{M.}~\bibnamefont{Schmidt}},
  \bibinfo{journal}{{Phys. Rev. E}} \textbf{\bibinfo{volume}{{81}}},
  \bibinfo{pages}{{041401}} (\bibinfo{year}{{2010}}).

\bibitem[{\citenamefont{Shundyak and van Roij}({2003})}]{shundyak_pre2003}
\bibinfo{author}{\bibfnamefont{K.}~\bibnamefont{Shundyak}} \bibnamefont{and}
  \bibinfo{author}{\bibfnamefont{R.}~\bibnamefont{van Roij}},
  \bibinfo{journal}{{Phys. Rev. E}} \textbf{\bibinfo{volume}{{68}}},
  \bibinfo{pages}{{061703}} (\bibinfo{year}{{2003}}).

\bibitem[{\citenamefont{Bier et~al.}({2004})\citenamefont{Bier, Harnau, and
  Dietrich}}]{bier_pre2004}
\bibinfo{author}{\bibfnamefont{M.}~\bibnamefont{Bier}},
  \bibinfo{author}{\bibfnamefont{L.}~\bibnamefont{Harnau}}, \bibnamefont{and}
  \bibinfo{author}{\bibfnamefont{S.}~\bibnamefont{Dietrich}},
  \bibinfo{journal}{{Phys. Rev. E}} \textbf{\bibinfo{volume}{{69}}},
  \bibinfo{pages}{{021506}} (\bibinfo{year}{{2004}}).

\bibitem[{\citenamefont{Vroege and
  Lekkerkerker}({1997})}]{vroege_lekkerkerker1997}
\bibinfo{author}{\bibfnamefont{G.}~\bibnamefont{Vroege}} \bibnamefont{and}
  \bibinfo{author}{\bibfnamefont{H.}~\bibnamefont{Lekkerkerker}},
  \bibinfo{journal}{{Colloids Surf., A}} \textbf{\bibinfo{volume}{{129}}},
  \bibinfo{pages}{{405}} (\bibinfo{year}{{1997}}).

\bibitem[{\citenamefont{Sollich et~al.}(2001)\citenamefont{Sollich, Warren, and
  Cates}}]{sollichwarrenmomenttheory}
\bibinfo{author}{\bibfnamefont{P.}~\bibnamefont{Sollich}},
  \bibinfo{author}{\bibfnamefont{P.~B.} \bibnamefont{Warren}},
  \bibnamefont{and} \bibinfo{author}{\bibfnamefont{M.~E.} \bibnamefont{Cates}},
  \bibinfo{journal}{Adv. Chem. Phys.} \textbf{\bibinfo{volume}{116}},
  \bibinfo{pages}{265} (\bibinfo{year}{2001}).

\bibitem[{\citenamefont{Sollich}(2002)}]{sollichoverview}
\bibinfo{author}{\bibfnamefont{P.}~\bibnamefont{Sollich}}, \bibinfo{journal}{J.
  Phys.; Condens. Matter} \textbf{\bibinfo{volume}{14}}, \bibinfo{pages}{79}
  (\bibinfo{year}{2002}).

\bibitem[{\citenamefont{Kofke and Bolhuis}(1999)}]{Kofke99}
\bibinfo{author}{\bibfnamefont{D.~A.} \bibnamefont{Kofke}} \bibnamefont{and}
  \bibinfo{author}{\bibfnamefont{P.~G.} \bibnamefont{Bolhuis}},
  \bibinfo{journal}{Phys. Rev. E} \textbf{\bibinfo{volume}{59}},
  \bibinfo{pages}{618} (\bibinfo{year}{1999}).

\bibitem[{\citenamefont{Wilding and Sollich}({2002})}]{wilding_jcp2002}
\bibinfo{author}{\bibfnamefont{N.}~\bibnamefont{Wilding}} \bibnamefont{and}
  \bibinfo{author}{\bibfnamefont{P.}~\bibnamefont{Sollich}},
  \bibinfo{journal}{{J. Chem. Phys.}} \textbf{\bibinfo{volume}{{116}}},
  \bibinfo{pages}{{7116}} (\bibinfo{year}{{2002}}).

\bibitem[{\citenamefont{Bartlett and Warren}({1999})}]{bartlett_prl1999}
\bibinfo{author}{\bibfnamefont{P.}~\bibnamefont{Bartlett}} \bibnamefont{and}
  \bibinfo{author}{\bibfnamefont{P.}~\bibnamefont{Warren}},
  \bibinfo{journal}{{Phys. Rev. Lett.}} \textbf{\bibinfo{volume}{{82}}},
  \bibinfo{pages}{{1979}} (\bibinfo{year}{{1999}}).

\bibitem[{\citenamefont{Phan et~al.}({1998})\citenamefont{Phan, Russel, Zhu,
  and Chaikin}}]{phan_jcp1998}
\bibinfo{author}{\bibfnamefont{S.}~\bibnamefont{Phan}},
  \bibinfo{author}{\bibfnamefont{W.}~\bibnamefont{Russel}},
  \bibinfo{author}{\bibfnamefont{J.}~\bibnamefont{Zhu}}, \bibnamefont{and}
  \bibinfo{author}{\bibfnamefont{P.}~\bibnamefont{Chaikin}},
  \bibinfo{journal}{{J. Chem. Phys.}} \textbf{\bibinfo{volume}{{108}}},
  \bibinfo{pages}{{9789}} (\bibinfo{year}{{1998}}).

\bibitem[{\citenamefont{Yiannourakou et~al.}({2009})\citenamefont{Yiannourakou,
  Economou, and Bitsanis}}]{yianno_jcp2009}
\bibinfo{author}{\bibfnamefont{M.}~\bibnamefont{Yiannourakou}},
  \bibinfo{author}{\bibfnamefont{I.~G.} \bibnamefont{Economou}},
  \bibnamefont{and} \bibinfo{author}{\bibfnamefont{I.~A.}
  \bibnamefont{Bitsanis}}, \bibinfo{journal}{{J. Chem. Phys.}}
  \textbf{\bibinfo{volume}{{130}}}, \bibinfo{pages}{{194902}}
  (\bibinfo{year}{{2009}}).

\bibitem[{\citenamefont{Sarkar et~al.}({2013})\citenamefont{Sarkar, Biswas,
  Santra, and Bagchi}}]{sarkar_pre2013}
\bibinfo{author}{\bibfnamefont{S.}~\bibnamefont{Sarkar}},
  \bibinfo{author}{\bibfnamefont{R.}~\bibnamefont{Biswas}},
  \bibinfo{author}{\bibfnamefont{M.}~\bibnamefont{Santra}}, \bibnamefont{and}
  \bibinfo{author}{\bibfnamefont{B.}~\bibnamefont{Bagchi}},
  \bibinfo{journal}{{Phys. Rev. E}} \textbf{\bibinfo{volume}{{88}}},
  \bibinfo{pages}{{022104}} (\bibinfo{year}{{2013}}).

\bibitem[{\citenamefont{van~der Linden et~al.}({2013})\citenamefont{van~der
  Linden, van Blaaderen, and Dijkstra}}]{vdlinden_jcp2013}
\bibinfo{author}{\bibfnamefont{M.~N.} \bibnamefont{van~der Linden}},
  \bibinfo{author}{\bibfnamefont{A.}~\bibnamefont{van Blaaderen}},
  \bibnamefont{and} \bibinfo{author}{\bibfnamefont{M.}~\bibnamefont{Dijkstra}},
  \bibinfo{journal}{{J. Chem. Phys.}} \textbf{\bibinfo{volume}{{138}}},
  \bibinfo{pages}{{114903}} (\bibinfo{year}{{2013}}).

\bibitem[{\citenamefont{Bartlett}({1998})}]{bartlett_jcp1998}
\bibinfo{author}{\bibfnamefont{P.}~\bibnamefont{Bartlett}},
  \bibinfo{journal}{{J. Chem. Phys.}} \textbf{\bibinfo{volume}{{109}}},
  \bibinfo{pages}{{10970}} (\bibinfo{year}{{1998}}).

\bibitem[{\citenamefont{Fasolo and Sollich}({2003})}]{fasolo_prl2003}
\bibinfo{author}{\bibfnamefont{M.}~\bibnamefont{Fasolo}} \bibnamefont{and}
  \bibinfo{author}{\bibfnamefont{P.}~\bibnamefont{Sollich}},
  \bibinfo{journal}{{Phys. Rev. Lett.}} \textbf{\bibinfo{volume}{{91}}},
  \bibinfo{pages}{{068301}} (\bibinfo{year}{{2003}}).

\bibitem[{\citenamefont{Sollich and Wilding}({2010})}]{sollich_prl2010}
\bibinfo{author}{\bibfnamefont{P.}~\bibnamefont{Sollich}} \bibnamefont{and}
  \bibinfo{author}{\bibfnamefont{N.~B.} \bibnamefont{Wilding}},
  \bibinfo{journal}{{Phys. Rev. Lett.}} \textbf{\bibinfo{volume}{{104}}},
  \bibinfo{pages}{{118302}} (\bibinfo{year}{{2010}}).

\bibitem[{\citenamefont{Botet et~al.}({2016})\citenamefont{Botet, Cabane,
  Goehring, Li, and Artzner}}]{botet_2016}
\bibinfo{author}{\bibfnamefont{R.}~\bibnamefont{Botet}},
  \bibinfo{author}{\bibfnamefont{B.}~\bibnamefont{Cabane}},
  \bibinfo{author}{\bibfnamefont{L.}~\bibnamefont{Goehring}},
  \bibinfo{author}{\bibfnamefont{J.}~\bibnamefont{Li}}, \bibnamefont{and}
  \bibinfo{author}{\bibfnamefont{F.}~\bibnamefont{Artzner}},
  \bibinfo{journal}{{Faraday Discuss.}} \textbf{\bibinfo{volume}{{186}}},
  \bibinfo{pages}{{229}} (\bibinfo{year}{{2016}}).

\bibitem[{\citenamefont{Cabane et~al.}({2016})\citenamefont{Cabane, Li,
  Artzner, Botet, Labbez, Bareigts, Sztucki, and Goehring}}]{cabane_prl2016}
\bibinfo{author}{\bibfnamefont{B.}~\bibnamefont{Cabane}},
  \bibinfo{author}{\bibfnamefont{J.}~\bibnamefont{Li}},
  \bibinfo{author}{\bibfnamefont{F.}~\bibnamefont{Artzner}},
  \bibinfo{author}{\bibfnamefont{R.}~\bibnamefont{Botet}},
  \bibinfo{author}{\bibfnamefont{C.}~\bibnamefont{Labbez}},
  \bibinfo{author}{\bibfnamefont{G.}~\bibnamefont{Bareigts}},
  \bibinfo{author}{\bibfnamefont{M.}~\bibnamefont{Sztucki}}, \bibnamefont{and}
  \bibinfo{author}{\bibfnamefont{L.}~\bibnamefont{Goehring}},
  \bibinfo{journal}{{Phys. Rev. Lett.}} \textbf{\bibinfo{volume}{{116}}}
  (\bibinfo{year}{{2016}}).

\bibitem[{\citenamefont{van~der Kooij et~al.}(2000)\citenamefont{van~der Kooij,
  Kassapidou, and Lekkerkerker}}]{kooij_nature2000}
\bibinfo{author}{\bibfnamefont{F.~M.} \bibnamefont{van~der Kooij}},
  \bibinfo{author}{\bibfnamefont{K.}~\bibnamefont{Kassapidou}},
  \bibnamefont{and} \bibinfo{author}{\bibfnamefont{H.~N.~W.}
  \bibnamefont{Lekkerkerker}}, \bibinfo{journal}{Nature}
  \textbf{\bibinfo{volume}{406}}, \bibinfo{pages}{868} (\bibinfo{year}{2000}).

\bibitem[{\citenamefont{Byelov et~al.}({2010})\citenamefont{Byelov, Mourad,
  Snigireva, Snigirev, Petukhov, and Lekkerkerker}}]{byelov_2010}
\bibinfo{author}{\bibfnamefont{D.~V.} \bibnamefont{Byelov}},
  \bibinfo{author}{\bibfnamefont{M.~C.~D.} \bibnamefont{Mourad}},
  \bibinfo{author}{\bibfnamefont{I.}~\bibnamefont{Snigireva}},
  \bibinfo{author}{\bibfnamefont{A.}~\bibnamefont{Snigirev}},
  \bibinfo{author}{\bibfnamefont{A.~V.} \bibnamefont{Petukhov}},
  \bibnamefont{and} \bibinfo{author}{\bibfnamefont{H.~N.~W.}
  \bibnamefont{Lekkerkerker}}, \bibinfo{journal}{{Langmuir}}
  \textbf{\bibinfo{volume}{{26}}}, \bibinfo{pages}{{6898}}
  (\bibinfo{year}{{2010}}).

\bibitem[{\citenamefont{Zhang and van Duijneveldt}({2006})}]{zhang_jcp2006}
\bibinfo{author}{\bibfnamefont{Z.}~\bibnamefont{Zhang}} \bibnamefont{and}
  \bibinfo{author}{\bibfnamefont{J.}~\bibnamefont{van Duijneveldt}},
  \bibinfo{journal}{{J. Chem. Phys.}} \textbf{\bibinfo{volume}{{124}}},
  \bibinfo{pages}{{154910}} (\bibinfo{year}{{2006}}).

\bibitem[{\citenamefont{Paineau et~al.}({2009})\citenamefont{Paineau, Antonova,
  Baravian, Bihannic, Davidson, Dozov, Imperor-Clerc, Levitz, Madsen, Meneau
  et~al.}}]{paineau_jpcb2009}
\bibinfo{author}{\bibfnamefont{E.}~\bibnamefont{Paineau}},
  \bibinfo{author}{\bibfnamefont{K.}~\bibnamefont{Antonova}},
  \bibinfo{author}{\bibfnamefont{C.}~\bibnamefont{Baravian}},
  \bibinfo{author}{\bibfnamefont{I.}~\bibnamefont{Bihannic}},
  \bibinfo{author}{\bibfnamefont{P.}~\bibnamefont{Davidson}},
  \bibinfo{author}{\bibfnamefont{I.}~\bibnamefont{Dozov}},
  \bibinfo{author}{\bibfnamefont{M.}~\bibnamefont{Imperor-Clerc}},
  \bibinfo{author}{\bibfnamefont{P.}~\bibnamefont{Levitz}},
  \bibinfo{author}{\bibfnamefont{A.}~\bibnamefont{Madsen}},
  \bibinfo{author}{\bibfnamefont{F.}~\bibnamefont{Meneau}},
  \bibnamefont{et~al.}, \bibinfo{journal}{{J. Phys. Chem. B}}
  \textbf{\bibinfo{volume}{{113}}}, \bibinfo{pages}{{15858}}
  (\bibinfo{year}{{2009}}).

\bibitem[{\citenamefont{Davidson et~al.}()\citenamefont{Davidson, Paineau, and
  Lekkerkerker}}]{pc1}
\bibinfo{author}{\bibfnamefont{P.}~\bibnamefont{Davidson}},
  \bibinfo{author}{\bibfnamefont{E.}~\bibnamefont{Paineau}}, \bibnamefont{and}
  \bibinfo{author}{\bibfnamefont{H.~N.~W.} \bibnamefont{Lekkerkerker}},
  \bibinfo{howpublished}{personal communication}.

\bibitem[{\citenamefont{Speranza and Sollich}(2003)}]{speranza_pre2003}
\bibinfo{author}{\bibfnamefont{A.}~\bibnamefont{Speranza}} \bibnamefont{and}
  \bibinfo{author}{\bibfnamefont{P.}~\bibnamefont{Sollich}},
  \bibinfo{journal}{Phys. Rev. E} \textbf{\bibinfo{volume}{67}},
  \bibinfo{pages}{061702} (\bibinfo{year}{2003}).

\bibitem[{\citenamefont{Wensink and Vroege}(2003)}]{wensink_jcp2003}
\bibinfo{author}{\bibfnamefont{H.~H.} \bibnamefont{Wensink}} \bibnamefont{and}
  \bibinfo{author}{\bibfnamefont{G.~J.} \bibnamefont{Vroege}},
  \bibinfo{journal}{J. Chem. Phys.} \textbf{\bibinfo{volume}{119}},
  \bibinfo{pages}{6868} (\bibinfo{year}{2003}).

\bibitem[{\citenamefont{Sollich}({2005})}]{sollich_jcp2005}
\bibinfo{author}{\bibfnamefont{P.}~\bibnamefont{Sollich}},
  \bibinfo{journal}{{J. Chem. Phys.}} \textbf{\bibinfo{volume}{{122}}},
  \bibinfo{pages}{{214911}} (\bibinfo{year}{{2005}}).

\bibitem[{\citenamefont{Onsager}(1949)}]{Onsager}
\bibinfo{author}{\bibfnamefont{L.}~\bibnamefont{Onsager}},
  \bibinfo{journal}{Ann. N.Y. Acad. Sci.} \textbf{\bibinfo{volume}{51}},
  \bibinfo{pages}{627} (\bibinfo{year}{1949}).

\bibitem[{\citenamefont{Odijk}(1986)}]{odijk1986}
\bibinfo{author}{\bibfnamefont{T.}~\bibnamefont{Odijk}}, \bibinfo{journal}{Liq.
  Cryst.} \textbf{\bibinfo{volume}{1}}, \bibinfo{pages}{97}
  (\bibinfo{year}{1986}).

\bibitem[{\citenamefont{Odijk and Lekkerkerker}(1985)}]{OdijkLekkerkerker}
\bibinfo{author}{\bibfnamefont{T.}~\bibnamefont{Odijk}} \bibnamefont{and}
  \bibinfo{author}{\bibfnamefont{H.~N.~W.} \bibnamefont{Lekkerkerker}},
  \bibinfo{journal}{J. Phys. Chem.} \textbf{\bibinfo{volume}{89}},
  \bibinfo{pages}{2090} (\bibinfo{year}{1985}).

\bibitem[{\citenamefont{Vroege and Lekkerkerker}(1993)}]{vroege1993}
\bibinfo{author}{\bibfnamefont{G.~J.} \bibnamefont{Vroege}} \bibnamefont{and}
  \bibinfo{author}{\bibfnamefont{H.~N.~W.} \bibnamefont{Lekkerkerker}},
  \bibinfo{journal}{J. Phys. Chem.} \textbf{\bibinfo{volume}{97}},
  \bibinfo{pages}{3601} (\bibinfo{year}{1993}).

\bibitem[{\citenamefont{Press et~al.}(1989)\citenamefont{Press, Teukolsky,
  Vetterling, and Flannery}}]{numrecipes}
\bibinfo{author}{\bibfnamefont{W.~H.} \bibnamefont{Press}},
  \bibinfo{author}{\bibfnamefont{S.~A.} \bibnamefont{Teukolsky}},
  \bibinfo{author}{\bibfnamefont{W.~T.} \bibnamefont{Vetterling}},
  \bibnamefont{and} \bibinfo{author}{\bibfnamefont{B.~P.}
  \bibnamefont{Flannery}}, \emph{\bibinfo{title}{Numerical Recipes}}
  (\bibinfo{publisher}{Cambridge University Press}, \bibinfo{address}{New
  York}, \bibinfo{year}{1989}).

\bibitem[{\citenamefont{Schulz}(1939)}]{schulz1939}
\bibinfo{author}{\bibfnamefont{G.~V.} \bibnamefont{Schulz}},
  \bibinfo{journal}{Z. Physik. Chem.} p.~\bibinfo{pages}{25}
  (\bibinfo{year}{1939}).

\bibitem[{\citenamefont{Zimm}(1948)}]{zimm1948}
\bibinfo{author}{\bibfnamefont{B.~H.} \bibnamefont{Zimm}}, \bibinfo{journal}{J.
  Chem. Phys.} \textbf{\bibinfo{volume}{16}}, \bibinfo{pages}{1099}
  (\bibinfo{year}{1948}).

\bibitem[{\citenamefont{Dhont}(1996)}]{dhontbook}
\bibinfo{author}{\bibfnamefont{J.~K.~G.} \bibnamefont{Dhont}},
  \emph{\bibinfo{title}{An Introduction to the Dynamics of Colloids}}
  (\bibinfo{publisher}{Elsevier}, \bibinfo{address}{Amsterdam},
  \bibinfo{year}{1996}).

\bibitem[{\citenamefont{Dhont and Briels}(2005)}]{dhontbrielsPRE2005}
\bibinfo{author}{\bibfnamefont{J.~K.~G.} \bibnamefont{Dhont}} \bibnamefont{and}
  \bibinfo{author}{\bibfnamefont{W.~J.} \bibnamefont{Briels}},
  \bibinfo{journal}{Phys. Rev. E} \textbf{\bibinfo{volume}{72}},
  \bibinfo{pages}{031404} (\bibinfo{year}{2005}).

\bibitem[{\citenamefont{Marconi and Tarazona}(1999)}]{marconi1999}
\bibinfo{author}{\bibfnamefont{U.~M.~B.} \bibnamefont{Marconi}}
  \bibnamefont{and} \bibinfo{author}{\bibfnamefont{P.}~\bibnamefont{Tarazona}},
  \bibinfo{journal}{J. Chem. Phys.} \textbf{\bibinfo{volume}{110}},
  \bibinfo{pages}{8032} (\bibinfo{year}{1999}).

\bibitem[{\citenamefont{Archer and Evans}(2004)}]{archer2004}
\bibinfo{author}{\bibfnamefont{A.~J.} \bibnamefont{Archer}} \bibnamefont{and}
  \bibinfo{author}{\bibfnamefont{R.}~\bibnamefont{Evans}}, \bibinfo{journal}{J.
  Chem. Phys.} \textbf{\bibinfo{volume}{121}}, \bibinfo{pages}{4246}
  (\bibinfo{year}{2004}).

\bibitem[{\citenamefont{Rex et~al.}(2007)\citenamefont{Rex, Wensink, and
  L\"{o}wen}}]{rex2007}
\bibinfo{author}{\bibfnamefont{M.}~\bibnamefont{Rex}},
  \bibinfo{author}{\bibfnamefont{H.~H.} \bibnamefont{Wensink}},
  \bibnamefont{and}
  \bibinfo{author}{\bibfnamefont{H.}~\bibnamefont{L\"{o}wen}},
  \bibinfo{journal}{Phys. Rev. E} \textbf{\bibinfo{volume}{76}},
  \bibinfo{pages}{021403} (\bibinfo{year}{2007}).

\end{thebibliography}

\end{document}